\documentstyle[pra,aps,twocolumn,epsfig]{revtex}

\begin{document}
\draft

\title{Quasi-periodic vs. irreversible dynamics of an optically confined
Bose-Einstein condensate }

\author{Pierre Villain, Patrik \"Ohberg and Maciej Lewenstein}

\address{Institut f\"ur Theoretische Physik, Universit\"at Hannover\\
Appelstr.2, 30167 Hannover, Germany}

\maketitle

\begin{abstract}
We consider the evolution of a dilute Bose-Einstein condensate in an 
optical trap formed by a
doughnut laser mode. By solving a one dimensional Gross-Pitaevskii equation
and looking at the variance and the statistical entropy associated with the
position of the system we can study the dynamical behavior of the system. It is
shown that for small condensates nonlinear revivals of the macroscopic wave
function are expected. For sufficiently large and dense
 condensates irreversible dynamics 
takes
place for which revivals of regular dynamics appear as predicted in
\cite{villainfpu}. These results are confirmed by a two dimensional simulation
in which the scales of energy associated with the two different directions
mimic the experimental situation.  
\end{abstract}


\section{Introduction}
The experimental achievement of Bose-Einstein condensation (BEC) in
cold alkali  atomic samples \cite{bec-Rb,bec-Na,bec-Li} has opened up  new
directions in atom optics. The  development of optical (laser) devices
to manipulate coherently atomic beams  has been particularly
spectacular. In the recent years 
it has led  to the advances in  atom cooling, providing thus the first steps
towards the realization of coherent matter waves. Atomic beam splitters, or
atomic mirrors are now common tools in quantum optics laboratories 
\cite{esslinger93,seifert94}. The
 Bose-Einstein condensates find  their natural place in this
research field as a source of intense coherent matter waves. Recently the
(coherent) bouncing of a BEC on an atomic mirror has been carefully studied
\cite{bouncing}. Pursuing the analogy with pure optical phenomena, a four wave
mixing experiment using BECs has been successfully realized \cite{phil}.
However, contrary to a standard four wave mixing of electromagnetic waves, 
the creation
of a fourth wave does not require any interaction with an external nonlinear
medium, but results directly from the intrinsic 
inter atomic interactions. This fact marks
a dramatic difference between the behavior of matter waves and light. 

For future applications of  those atomic coherent matter waves, 
it is of great importance to study in detail  the role of the
interactions in the dynamics of the condensate. Such studies are of
 crucial interest
especially when using atomic waveguides and
atomic resonators. The main element of such atom optics tools is a far off
resonant hollow laser beam which can be loaded with a Bose-Einstein condensate
from an atomic trap \cite{Hann}. For a sufficiently large detuning,
spontaneous emission may be neglected leading to a coherent manipulation
of the motion of the atoms. If used alone, this doughnut laser beam may
be viewed as a matter wave guide, see Fig. 1. 
If two other blue detuned lasers are added
transversally, an atomic de Broglie cavity can be formed. 
\begin{figure}[ht]  
\begin{center} 
\includegraphics[width=6cm]{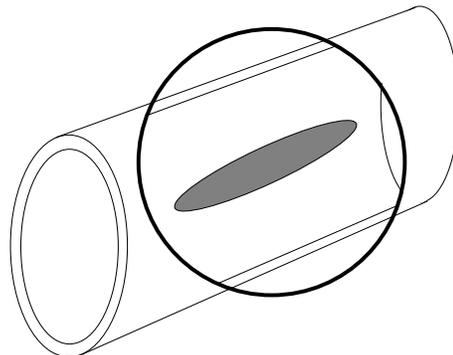}
\label{donut}
\caption{A schematic view of the condensate in the optical trap.}
\end{center}  
\end{figure}

In a previous paper \cite{villainfpu} we studied the dynamics of a condensate
inside a simple resonator (modelled by a one dimensional square box) and
focused particularly on the influence of the inter atomic interactions on the
behavior of the system. We showed that starting from a localized wave packet
and by increasing the nonlinearities we pass from a quasi linear regime, where
phenomena like nonlinear revivals and fractional revivals are expected
\cite{jane}, to a regime where stochastization of the movement takes place and
where the nonlinear revivals of the wave packet are replaced by  revivals of
regular dynamics. 

In this paper we are interested in finding the same kind of phenomena but 
using a more
realistic trapping potential. In particular
the rigid boundary conditions corresponding to 
 the square box potential have to
be modified, since this kind of discontinuities cannot be experimentally
realized. Moreover, in order to probe the dynamics of the system, some
experimentally accessible tools, more adequate than those used in
\cite{villainfpu}, have to be introduced.
It must be stressed that the (nonlinear) revivals investigated in this paper
concern a mesoscopic object following a {\it 
nonlinear evolution equation}.
So far the revivals of a wave packet have been studied
\cite{averbukh,eberly}, and observed only for {\it microscopic} and {\it 
linear} systems
\cite{stroud,ewart,stolow,wineland,haroche}. On the other hand, the
irreversible behavior which may appear in the dynamics, 
concerns the evolution of an isolated
system and is only due to the nonlinear  character of the evolution; it is 
not caused by
any damping originating from some interactions with an external reservoir.

The plan of the paper is the following: in section II, after describing
briefly the model which we use, we introduce some
experimentally accessible observables, namely the variance and the entropy
associated with the position, which will be used to check  the dynamical   
behavior of the system. We will show how to recover the results of Ref.
\cite{villainfpu} (obtained for a box potential)
in the cases most interesting for experiments. In section III
we use the observables of section II in the case of a more 
realistic potential. We discuss how the 
different regimes of the condensate behavior is reflected in the 
behavior of the variance and the entropy. In section IV, in order to
support  the validity of the one dimensional model we present some results
obtained in a 2D simulation.  Finally, we conclude in Sec. V. 

\section{Model and new variables}

We recall here briefly the mathematical model we use in order to describe the
evolution of a Bose-Einstein condensate in an atomic cavity.
We consider the zero temperature case and use the standard Hartree-Bogoliubov
approach \cite{nozieres}. All the atoms are described by the same wave
function $\Psi(\vec r, t)$, the evolution of which is given by the time
dependent Gross-Pitaevskii equation
\begin{eqnarray}
i\hbar\frac{\partial \Psi(\vec r,t)}{\partial
t}&=&\left[-\frac{\hbar^2}{2m}\nabla^2+ V_{trap}(\vec r,t)\right]\Psi(\vec
r,t)\nonumber\\  &+&Nu_0|\Psi(\vec r,t)|^2\Psi(\vec r,t).\label{GP}
\end{eqnarray}
We denote by $u_0 =4\pi\hbar^2 a_{\rm
sc}/m$, with $a_{sc}$ being the $s$-wave scattering length, and by $N$ the total
number of atoms. 
The core of the atomic resonator we have in mind is a Laguerre-Gauss laser
mode. The radial confinement due to this optical potential is very strong
compared to the longitudinal one. Typically the internal radius of the
"cylinder" is of the order of a few micrometers and the length $L$ of the 
"box" may vary from 75 to 100 micrometers. This leads to a radial
confinement of the order of several hundreds Hz, while  the longitudinal scale
of frequency  will be of the order of  few Hz. 
Unless we use a very large number of atoms, the
dynamics in the radial direction can be considered as frozen; the system 
behaves effectively as a 1D one, 
relevant dynamics occurs then
   in the longitudinal direction (we will come back to
this assumption in Sec. IV). Therefore we use a wave function in a 
product form 
$\Psi(\vec r,t)= \psi(z,t) \chi_{0}(x,y,t)$, with $\chi_0$ denoting the
ground state
in the transverse direction. We simplify now the trapping potential by using
a square box potential.  The evolution equation for $\psi(z,t)$
is then 
\begin{eqnarray}  
i\hbar\frac{\partial \psi(z,t)}{\partial
  t}=-\frac{\hbar^2}{2m}\frac{\partial^2\psi(z,t)}{\partial z^2}  +\frac{Nu_0}{
S}|\psi(z,t)|^2\psi(z,t),
\end{eqnarray} 
with $\psi$ fulfilling rigid boundary conditions, and  $S$ denoting the
(effective) transversal section that we assume to be 
constant over the total length of the box.
This is of course an additional approximation, since the doughnut laser beam 
is always
focused in the region where the loading of the condensate takes place.
Consequently, rigorously speaking we have $S=S(z)$ 
with $S(z)$ exhibiting a minimum in the
center, and a maximum at the borders. Since the relative variance of 
$S(z)$ is rather small
\cite{hannoverlaser}, we may neglect  it in the first approximation.  

Depending on the value of the nonlinear parameter, the
dynamics described by Eq.(2) can attain very different character
\cite{villainfpu}. 
The relevant
nonlinear parameter is given by the ratio $\frac{\Omega_{int}}{\omega_1}$
where $\Omega_{int}= Nu_0/\hbar$ is the characteristic frequency related to the
interaction term and $\omega_1$ is the frequency of the fundamental mode
$\omega_1= \frac{\pi^2 \hbar}{2mL^2}$. We investigate the
dynamical behavior for the values $\frac{\Omega_{int}}{\omega_1} =3.08$ and 
$\frac{\Omega_{int}}{\omega_1}=61$ which correspond to a typical number of
atoms from several hundreds to  few thousands in the first case, and from  few
thousands to  few tens of thousands in the second one. Independently  of the
chosen initial state, we observe  that for 
$\frac{\Omega_{int}}{\omega_1}=3.08$ 
the dynamics remains regular and quasi-periodic (nonlinear
revivals of the wave function occur), whereas for 
$\frac{\Omega_{int}}{\omega_1}=61$ the
motion is stochastic. In the latter case the wave function does 
not exhibit any
strong spatial relocalization, but the previously mentioned
 nonlinear revivals  are  
replaced now by {\it revivals of regular dynamics}
 characterized by a return of
most of the energy in the few initially populated modes. These results
were obtained in Ref. \cite{villainfpu}
by checking the appearance of energy transfer between the
eigenmodes of the system and by looking at the temporal spectrum of the
density. 

Experimentally the access to the quantities discussed in Ref. 
\cite{villainfpu} is 
difficult, if not hardly possible. 
The simplest quantity to measure experimentally is the
local density $N|\psi(z,t)|^2$. In order to check the appearance of revivals of
the wave function or revivals of regular dynamics the most natural time 
dependent quantity
to study is the variance $\sigma_z$ of the mean position 
of the condensate cloud, defined as 
$\sigma^2_z= \langle(z-\langle z \rangle_{\psi})^2\rangle_{\psi}$ where
$\langle...\rangle_{\psi}= \int dz ...|\psi(z,t)|^2$. 
If this function exhibits a return to its initial value we will be
able to conclude about the appearance of nonlinear revivals of the wave
function, or equivalently to some spatial relocalization to its initial
position.
Although this quantity is adequate in order to probe the revivals, it is
insufficient to conclude without ambiguity that a complete delocalization 
of the wave function has occurred. Such delocalization may
 correspond either  to a relaxation 
into another regular state with
 a density peaked only at the cavity ends, or to  irreversible dynamics
leading to a more or less uniform density distribution. 
 We need 
therefore
another dynamical quantity which would be 
more sensitive to various  kinds of spatial
delocalization and relocalization of the wave packet. 
The entropy $S_z$ associated with the
position of the system fulfills such a requirement \cite{bialinicky}. This
statistical entropy is defined as 
$$S_z(t)= -\sum_n P_n(t) {\rm log} P_n(t),$$
with  $P_n$ being the probability of presence in the interval $[z_n,z_n+dz]$ 
contained in the
interval  $[0,L]$,
\begin{eqnarray} 
P_n(t) = |\psi(z_n, t)|^2 dz.
\end{eqnarray}
This function reaches its maximum value $S_z^{\rm max}$ 
for a uniform probability
distribution corresponding to the case of a spatially uniform system. On the
other hand any spatial localization of the wave function will induce a strong
decrease of $S_z$ compared to the maximum value.  
As usual it is better not to use $S_z$ but a normalized
entropy $\eta_z$ defined as
\begin{eqnarray}
\eta_z(t) = \frac{S_z^{\rm max}-S_z(t)}{S_z^{\rm max}-S_z(0)}.\label{entropy}
\end{eqnarray}

Experimentally, both $\sigma_z$ and
$S_z$ correspond to measurable quantities. The only restriction is coming from
the (inevitably) limited spatial resolution in a given experiment. But this
limitation affects only the dynamics associated with high frequency modes and
the interesting phenomena (revivals of the wave packet or revivals of regular
dynamics) are in fact associated to the low modes of
the system. A typical spatial resolution can be of the order 3$\mu$m which, 
in a box of 100$\mu$m limits the resolution
of the modes higher than the 30th. 

We simulate the evolution of a wave packet initially centered in the box with
no initial velocity and with an initial  width of $L/5$. This
corresponds to the first case investigated in \cite{villainfpu} where 
mainly the modes 1
and 3 of the box potential are populated. We use the usual split
operator method to solve Eq.(2) \cite{numerical}. We check the conservation
of the norm of the wave function and the conservation of the total energy. The
former was conserved better than $10^{-4}\%$, and the latter has shown
fluctuations smaller than $1\%$.  In the figures 2a and 2b we present
respectively the time evolution of the variance compared to its value for a
uniform system $\sigma_{uni}$ and the time evolution of $\eta_z$. These
plots correspond to a nonlinear coefficient
$\frac{\Omega_{int}}{\omega_1} =3.08$. The time is given in units of the
fundamental period of the box potential $T_1= 2\pi/\omega_1$. We restrict 
ourself to a time interval of one period roughly since for the
experimental system  it would correspond to a time of the order of 18s  which
is beyond, or in the best case close to the usual lifetime of 
the condensate in this situation. 

\begin{figure}[ht]  
\begin{center} 
\includegraphics[width=6cm]{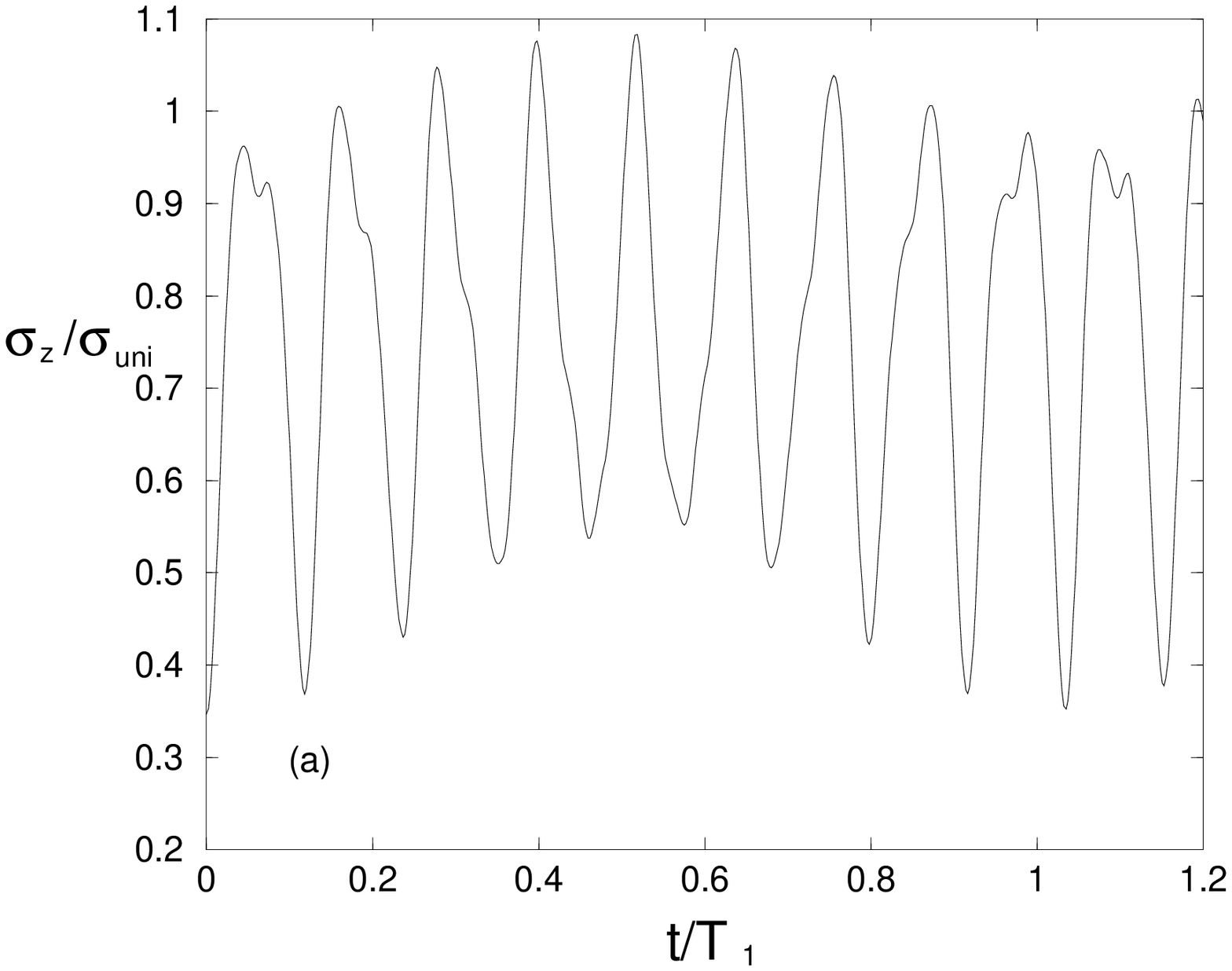}\\
\includegraphics[width=6cm]{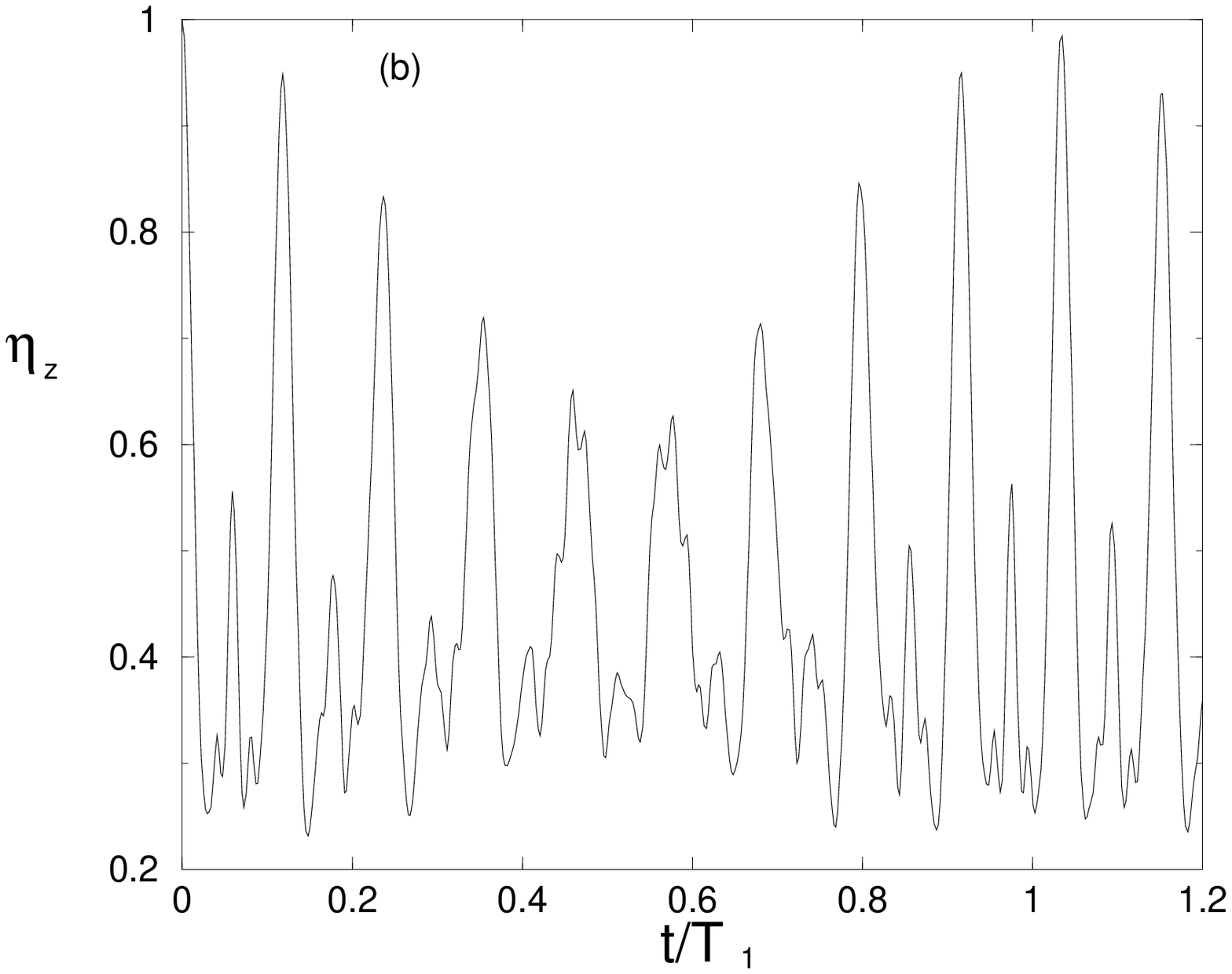}
\label{Fig.1}
\caption{Evolution of (a) $\sigma_z/\sigma_{uni}$ and (b) 
$\eta_z$ in the case $\frac{\Omega_{int}}{\omega_1}=3.08$ and for an
initially centered wave packet.} 
\end{center}  
\end{figure}

The evolution of $\sigma_z$ shows clearly some quasi-periodic revivals at
times close to those expected in the linear case with this initial
state. The nonlinear character of the evolution appears through the position
of this revival. The behavior of $\eta_z$ confirms the
regular spatial relocalization of the wave function. This function exhibits
clear returns to its maximal value 1. Notice that the fractional revivals  
corresponding to a localization of the atoms near the boundaries are not
at all exhibited  by $\sigma_z$, but appear very clearly in the time 
dependence of  $\eta_z$, as
expected. They are represented by the small peaks preceding a quasi-revival of the
system. The occurrence of quasi periodic nonlinear revivals of a mesoscopic 
wave packet is thus confirmed. From the experimental point of view
the bad point is of course the time scale of these phenomena.  The
first nonlinear revival is shown at a time close to $0.1 T_1$ ($\simeq 1.8$s),
 which is already
rather long. We will see in section III that in a more realistic potential
than a box like potential 
the revival time is significantly reduced.

In the figures 3a and 3b we present the time evolution of the same quantities
but for a nonlinear parameter  $\frac{\Omega_{int}}{\omega_1}=61$.
\begin{figure}[ht] 
\begin{center} 
\includegraphics[width=6cm]{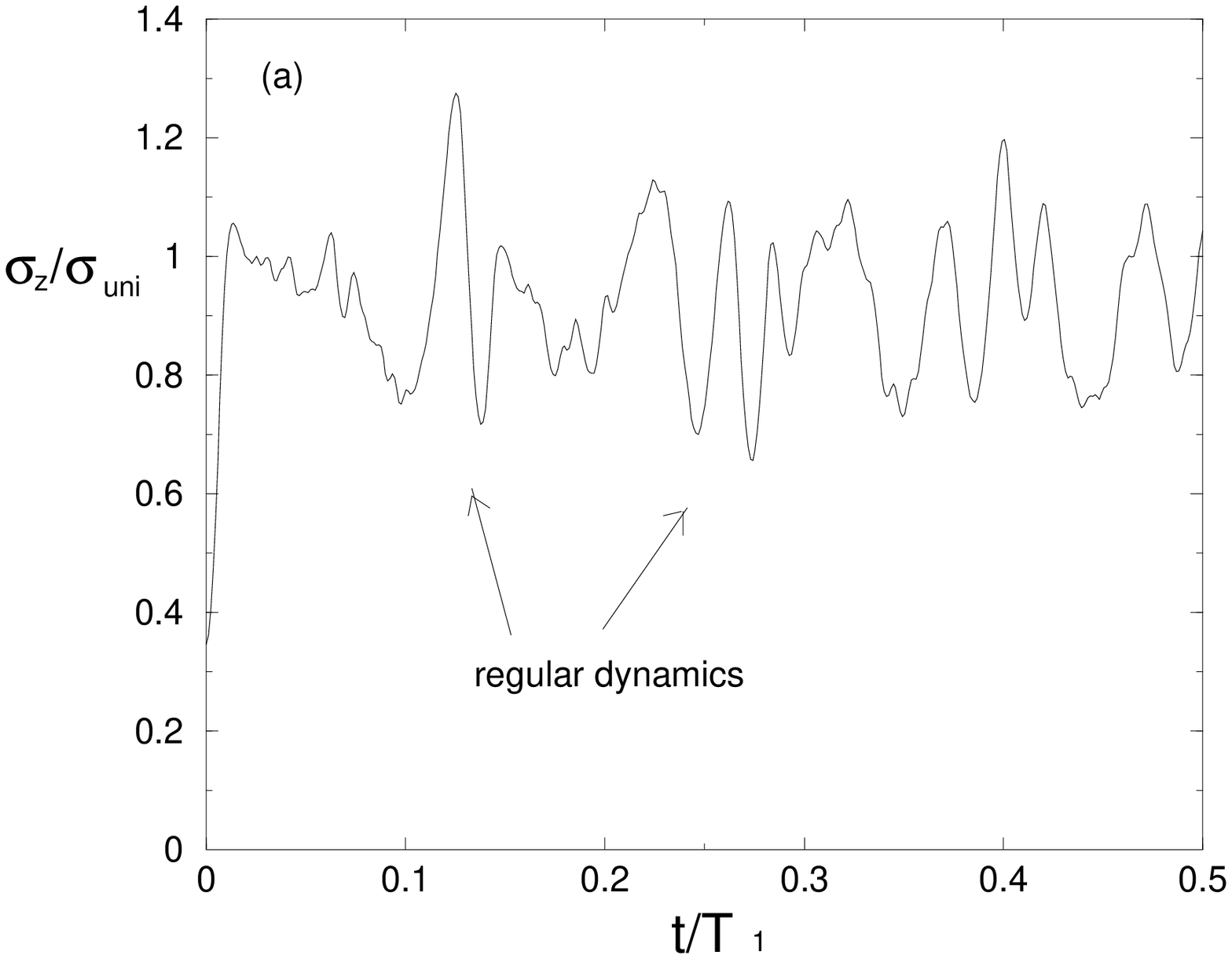}\\
\includegraphics[width=6cm]{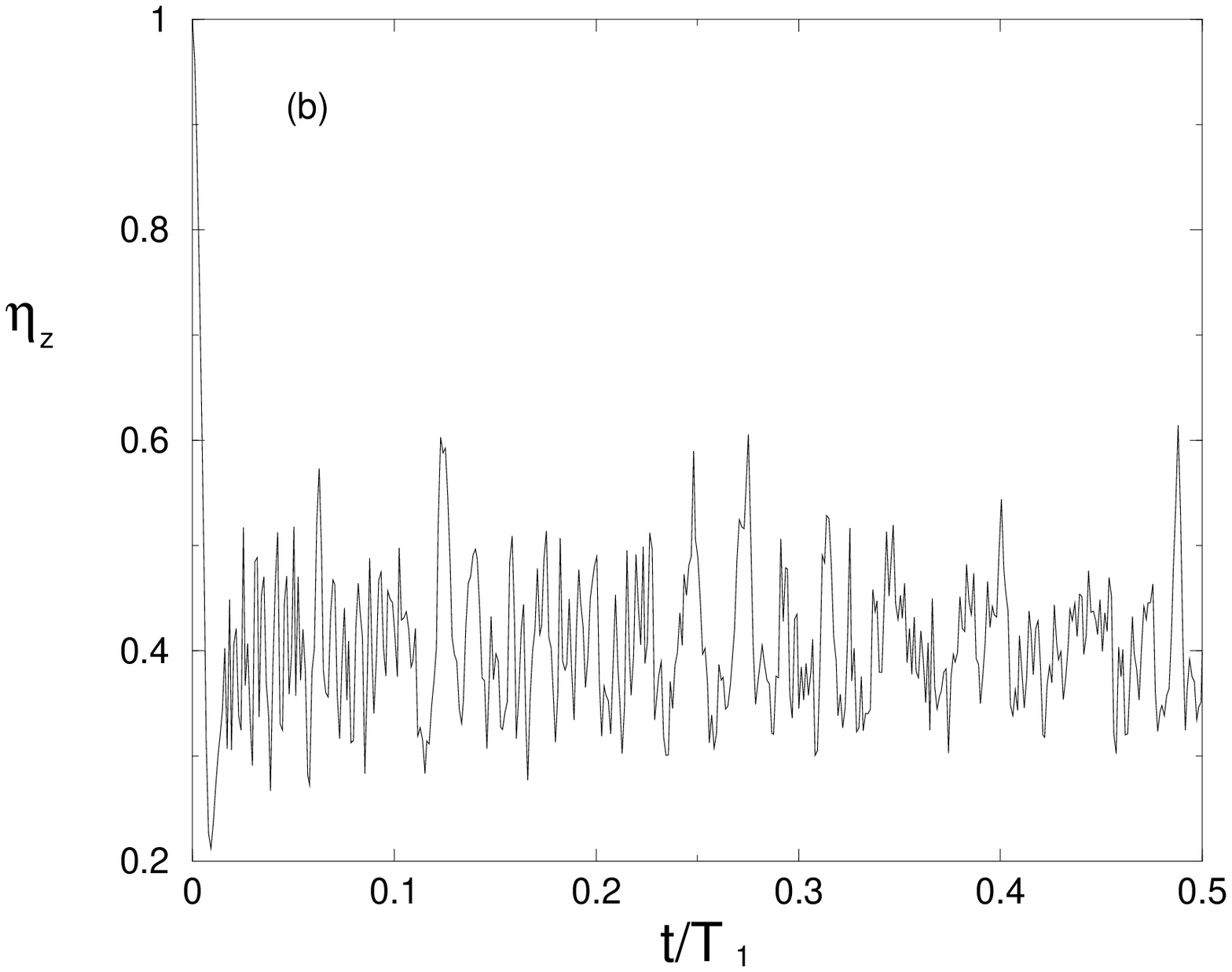}
\label{Fig. 2}
\caption{Evolution of (a) $\sigma_z/\sigma_{uni}$ and (b) 
$\eta_z$ in the case $\frac{\Omega_{int}}{\omega_1}=61$ and for an
initially centered wave packet.}
\end{center}  
\end{figure}

The variance does not  exhibit any clear return to its initial value. On the
contrary it oscillates around the value $\sigma_{uni}$. The  absence of strong
spatial relocalizations of the wave function is confirmed by the behavior of
$\eta_z$ which stays far from 1 and oscillates around a low value.
Notice that this value is not zero, which means that the system has
effectively relaxed but is still following some dynamics. The change from
the nonlinear revivals of $\psi$ to the revivals of regular dynamics is
however visible. The occurrence of regular dynamics during some time intervals
is indeed clearly exhibited by the variance. At these times where the energy
of the system comes back to the low modes of the box potential the variations 
of $\sigma_z$  
have a significant lower rate as shown in Fig. 3a. This kind of behavior
appears also in the variations  of $\eta_z$ but is not as spectacular as for
$\sigma_z$.

At this point the variance and the entropy associated with the
position appear to be good physical quantities to probe the dynamical
behavior of the system and for showing its particularities (revivals of the
wave packet or of regular dynamics). However, the time scale 
needed for observation of these phenomena 
is too long from the experimental point of view. 
This fact is mainly due to the form of the chosen potential. We investigate
in the next section how these results are modified by introducing 
more realistic boundary conditions.

\section{Studies of the dynamics in a realistic trapping potential}

Experimentally the optical box will be closed transversally by using two 
blue detuned lasers. The trapping potential at the boundaries of the box is no
more discontinuous and may be modelled in a good approximation as the sum 
 of two  identical Gaussians centered at $z=0$ and $z=L$.   
The characteristic parameters of the potential are then its height $V_{laser}$
and the width of the Gaussian $\sigma_{laser}$. The requirement on
$V_{laser}$ is that it provides a sufficiently high barrier in order to
prevent the atoms from leaving the trap. We take it as a given parameter of the
problem. The value of $\sigma_{laser}$ directly depends
on the transversal focusing of the lasers at $z=0$ and $z=L$. The
typical value cannot be smaller than a few $\mu m$. 
Clearly a too small $\sigma_{laser}$ will take us back to the
case of the perfect box.  It is beneficial to use a large $\sigma_{laser}$ 
in order to reduce the time scale on which the
nonlinear revivals of the wave packet or of the regular dynamics appear.
On the other hand, too large $\sigma_{laser}$ does not interest us  because
it will correspond to a situation, where the atoms are still very confined
and the cloud will just oscillate as in a slightly modified harmonic trap.
In other words, the overlapping of the boundary barriers with the initial
state must remain as small as possible in order to remain in a situation
similar to a box potential from the  dynamical point of view.  

The study of the dynamics of the system appears to be simpler if one uses
the quantities $\eta_z$ and $\sigma_z$, since it is difficult to have access
to a large range of different modes of the potential (which would be necessary
in order to compute for instance the statistical entropy associated with the
energy distribution among the eigenmodes as in Ref. \cite{villainfpu}).
We present here the results of the simulations for the same case as before, namely 
an initially centered wave packet with a zero mean velocity for the two same 
values of the nonlinear parameter. The width of the wave packet is still 
$L/5$ and we choose a $\sigma_{laser}= L/12$ which corresponds in a typical 
experiment to a
focused waist of roughly $16\mu$m and the height of the barrier is chosen to
be of the order a few mK.
We show the time variations of the same
quantities than before in order to compare the results with the case of the 
perfect box. The variance $\sigma_z/\sigma_{uni}$ is still 
expressed in units of 
the uniform case which is calculated using the box potential. 
We keep also the same unit of time which will allow
us to compare the time scale obtained now for the different phenomena we want
to investigate. In Figs. 4a and 4b we show the variations of the variance and
the normalized entropy $\eta_z$ for $\frac{\Omega_{int}}{\omega_1}=3.08$.

\begin{figure}[ht] 
\begin{center} 

\includegraphics[width=6cm]{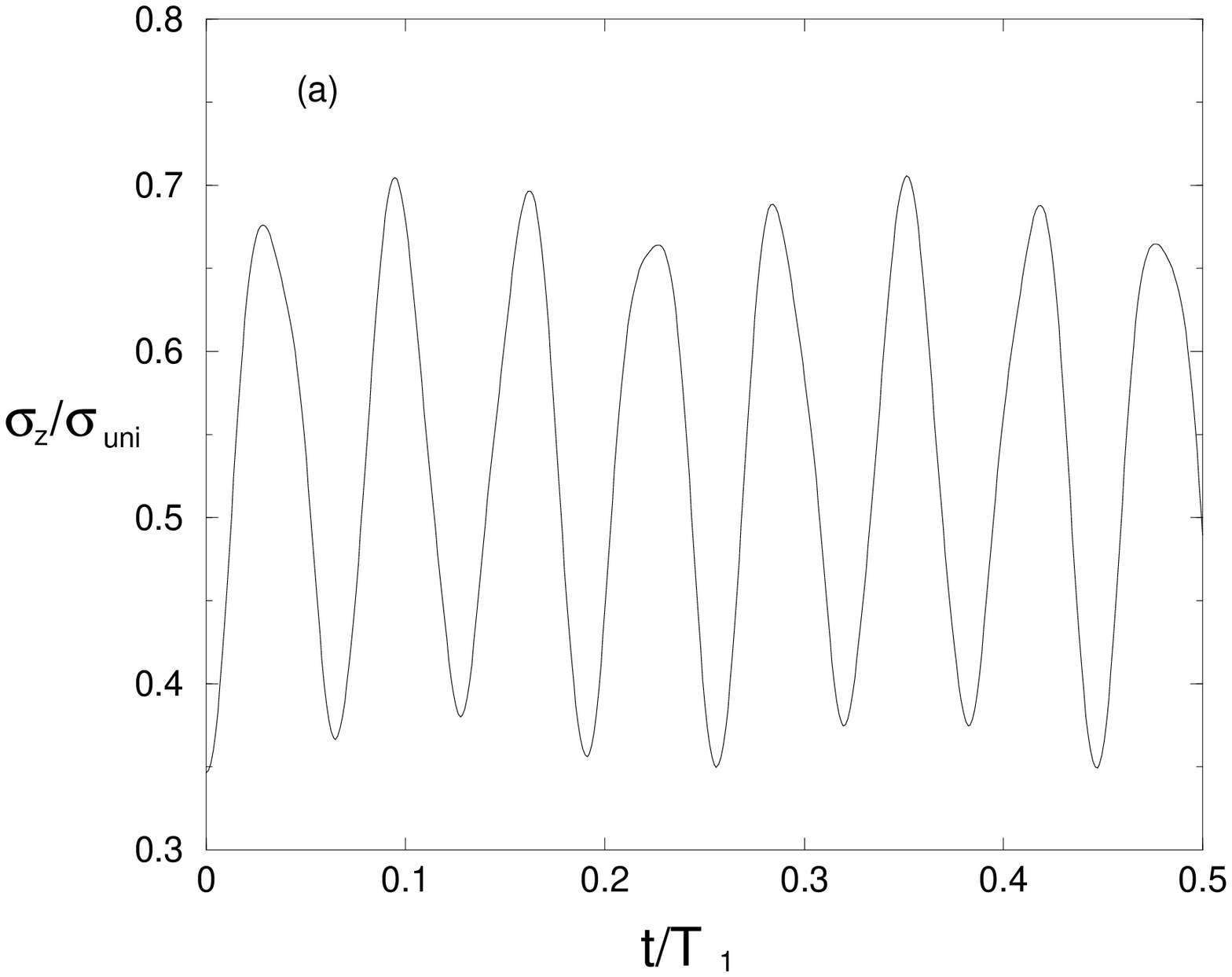}\\
\includegraphics[width=6cm]{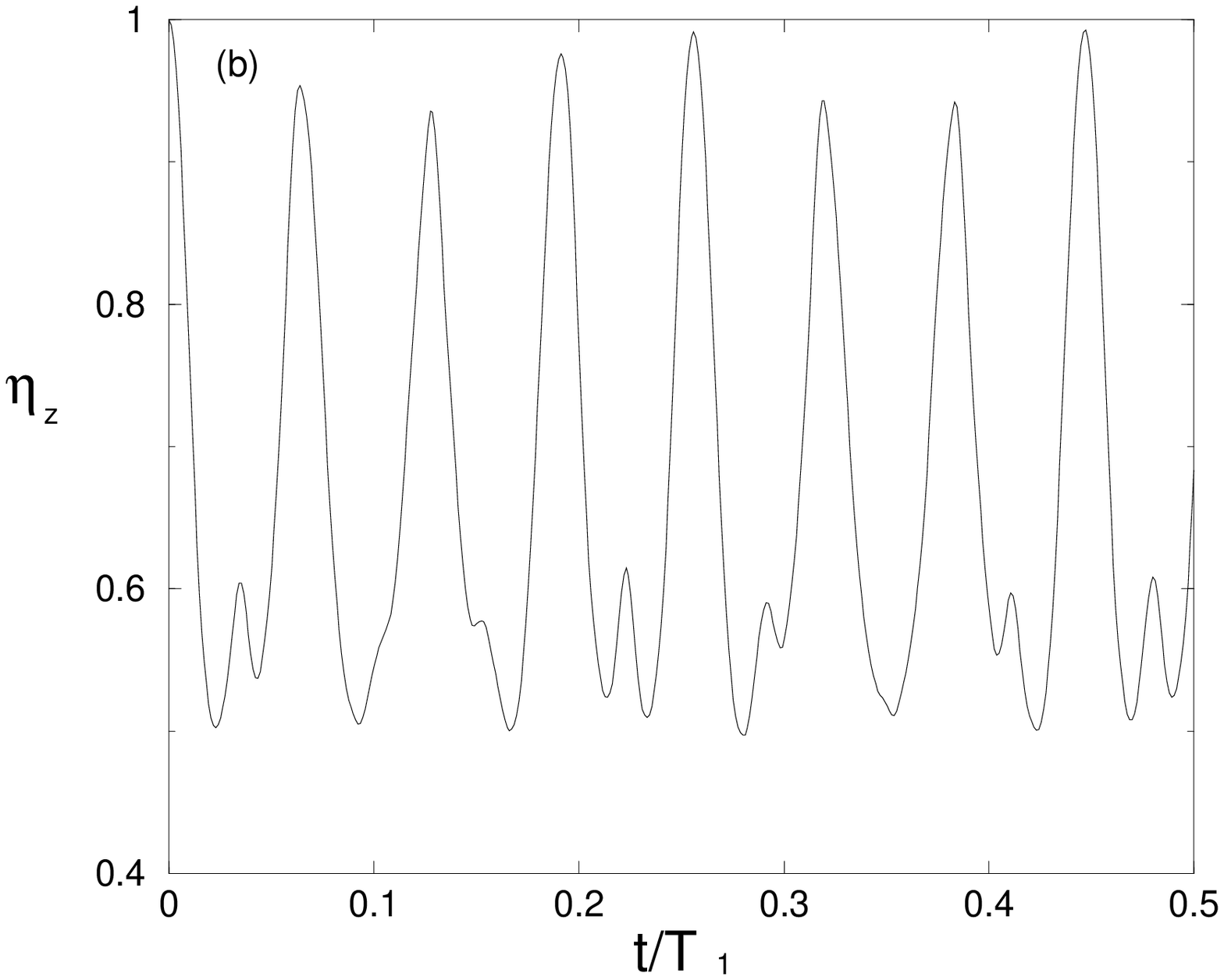}
\label{Fig. 3}

\caption{ Temporal variation of (a) $\sigma_z/\sigma_{uni}$ and (b) 
$\eta_z$ for an initial centered state without mean velocity
where the boundary potential has a Gaussian shape and 
$\frac{\Omega_{int}}{\omega_1}=3.08$}
 \end{center}  
\end{figure}

The regular and quasi-periodic behavior of the two quantities is striking.
The wave packet clearly exhibits nonlinear revivals at almost regular times.
Although the maximal value of $\sigma_z$ is as expected 
smaller than for the perfect box, 
the most important point is that the time scale
for the nonlinear revivals has been divided by a factor 2. This makes the
first quasi-revival of the condensate to appear before 1s. Although it
remains long, it is now sufficiently shorter than the lifetime of the 
condensate. 
This time scale can be decreased even more by playing with the value
of $\sigma_{laser}$ as well as with the initial state of the system. In
particular a spatially more squeezed initial wave packet will populate
higher excited states at the beginning of the evolution 
leading to a smaller linear
revival time, and, consequently, to a smaller nonlinear 
revival time (with these
values of the nonlinearities the two regimes are not too much different
and the time scales are comparable). It is important to note that, as before, 
some fractional
revivals corresponding to an accumulation of atoms at the boundaries are still
visible in the variations of $\eta_z$.

We turn now to the case of strong nonlinearities.
The simulations of the evolution $\sigma_z$ and $\eta_z$ for a nonlinear
parameter  $\frac{\Omega_{int}}{\omega_1}=61$ are presented in Figs. 
5a and 5b keeping the same conventions as before.

\begin{figure}[ht] 
\begin{center} 
\includegraphics[width=6cm]{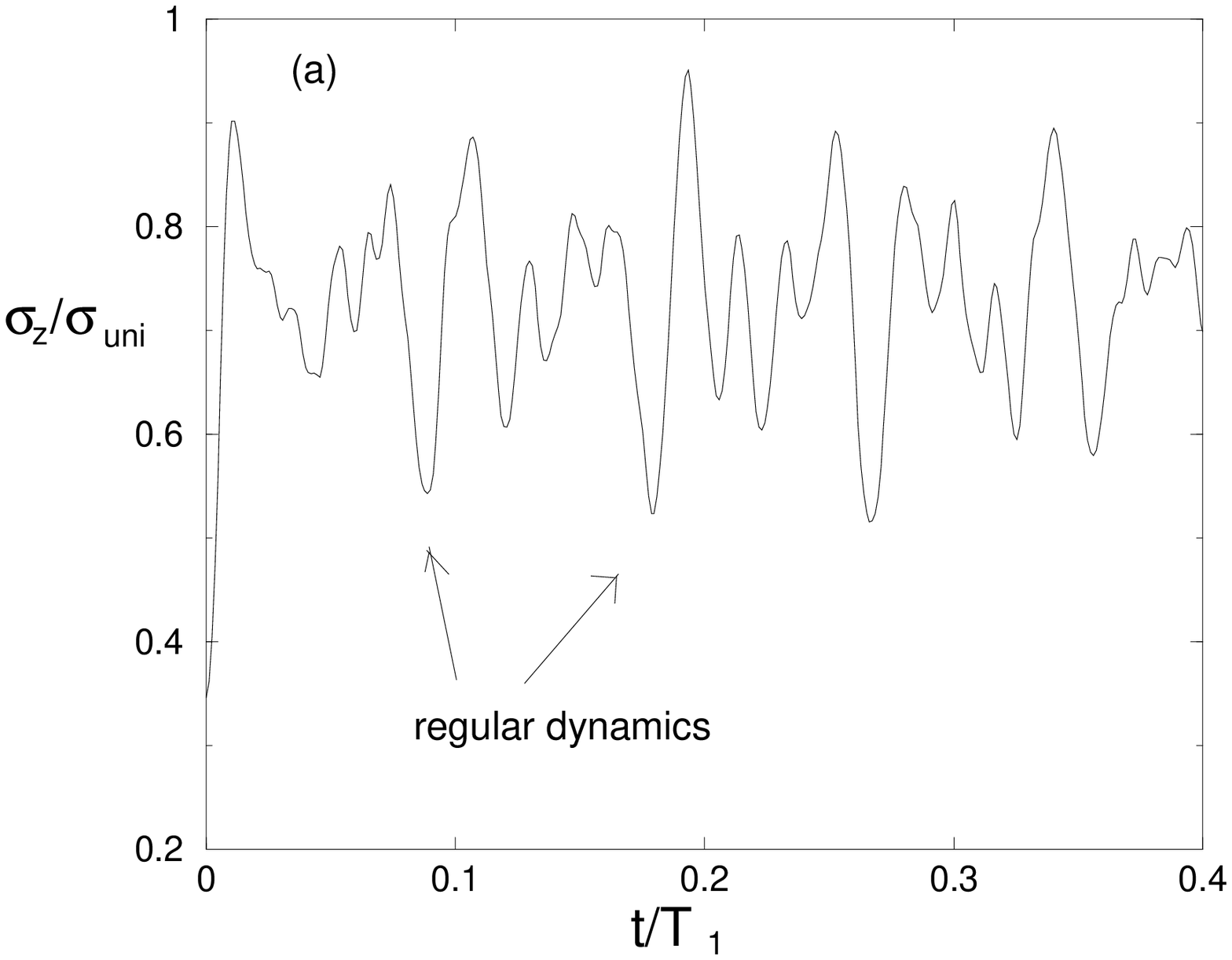} \\
\includegraphics[width=6cm]{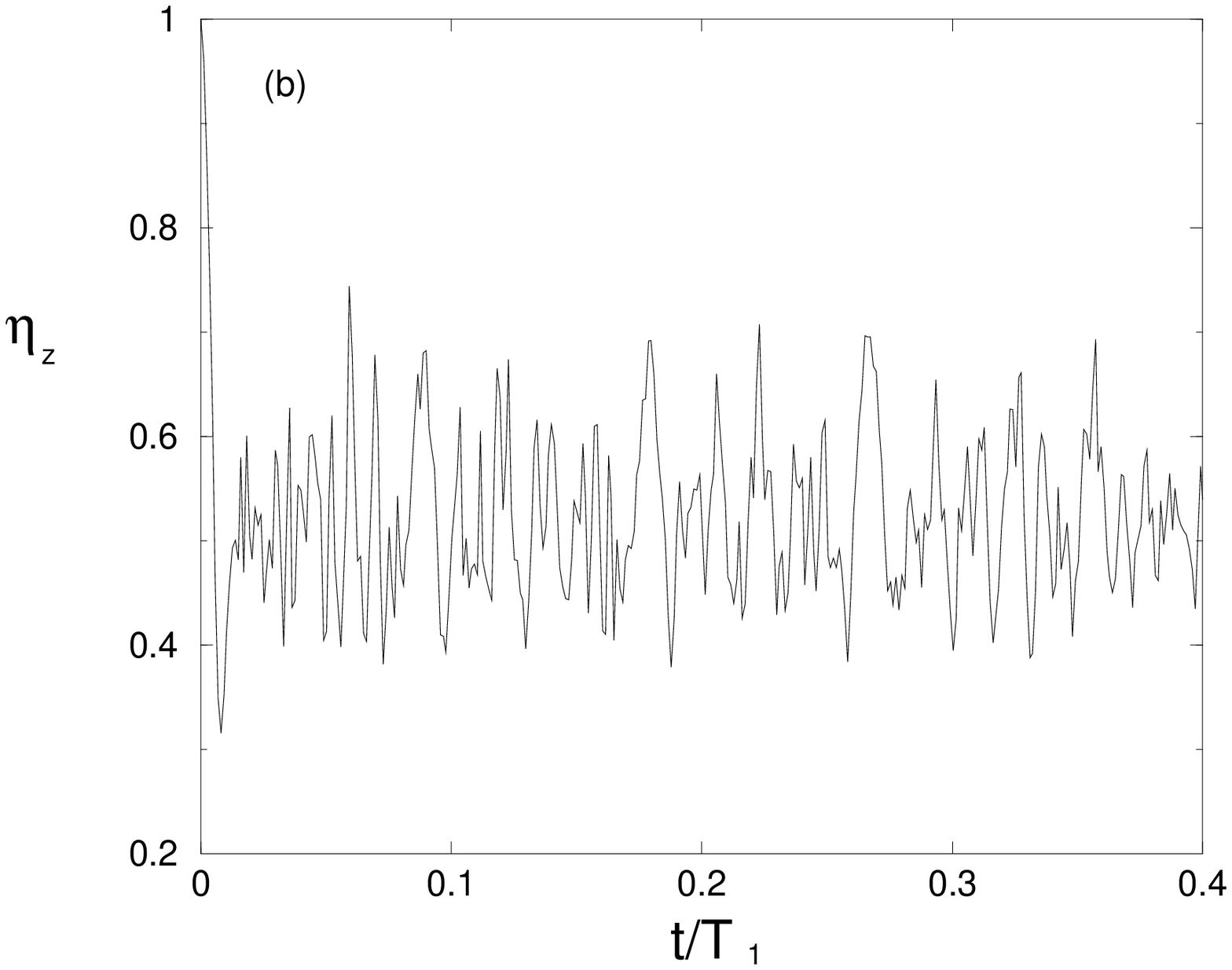}
\caption{Temporal variation of (a) $\sigma_z/\sigma_{uni}$ and (b) 
$\eta_z$ for an initial centered state without mean velocity
where the boundary potential has a Gaussian shape and
$\frac{\Omega_{int}}{\omega_1}=61$} 
\end{center}   
\end{figure}

The behavior is similar to those of Fig. 3. The variance does not come back
to its initial value and $\eta_z$ stays far from 1 which proves that the
wave function does not exhibit any spatial relocalization. The variations of
$\sigma_z$ still exhibit the revivals of regular dynamics. Although the time
scale on which they appear has been reduced it is still of the order of
$0.1T_1$ which makes them difficult to observe experimentally.

\section{Analysis of the two dimensional problem.}

\subsection{Analysis of the model.}

 In the previous 
sections we 
have studies the simplified  1D dynamics arguing 
that the  radial confinement due to the
doughnut laser mode was much stronger than the longitudinal one. In
consequence the structure of the condensate in the radial direction will not
be modified by the nonlinearities. Although this hypothesis is 
reasonable, it has to be checked at the level of the physical quantities we
are using to describe the behavior of the system. The
 ratio of the typical energies associated with the radial degrees of freedom
and the longitudinal ones is of the order of  $L^2/R^2 \simeq 100$ where
$R$ denotes the internal mean radius of the Laguerre-Gauss laser mode and
$L$ the length in the longitudinal direction.  Qualitatively, for small
nonlinearities, when  the dynamics is already quasi-periodic in
 the longitudinal
direction, there are no serious worries concerning the influence of the other
degrees of freedom. On the other hand, for stronger nonlinearities
corresponding to $\frac{\Omega_{int}}{\omega_1}$ of the same order of
magnitude as $L^2/R^2$, a strong coupling between the different degrees of
freedom associated with different directions may appear, and it is not sure
that we can get the same behavior as before for the variance and the entropy
associated with the longitudinal direction. However, we must keep in mind that
this simple energetic estimate does not take into account the population of
the modes which are important parameters in nonlinear problems. The derivation
of more accurate criteria is then necessary. We will propose some of them in
the next paragraph.
   
We turn now to the discussion of the inclusion of an additional dimension.
For this purpose we consider the evolution  of a condensate  
in a two-dimensional box. We denote the directions by $z$ and $y$ and by $L_z$
and $L_y$ the associated lengths. We take $L_z=4L_y$ where
the energy scale in the $y$ direction is 
$16$ times bigger than the one  of the $z$ direction and will mimic our
radial confinement in the real system. Of course as we are working in Cartesian
coordinates the problem is simplified compared to the cylindrical one but it
should give us a good representation of the real system. Moreover the typical
ratio of the energies is one order of magnitude smaller than the one of our
system but it will still provide the same kind of behavior and allows us to
test values of nonlinearities sufficiently small to have a good spatial
resolution in our numerical grid.

The starting point is still the time dependent Gross-Pitaevskii equation
(\ref{GP}). Assuming a very strong confinement in the $x$ direction, we
can neglect the degrees of freedom associated with the $x$ direction by
assuming a wave function of the form $\Psi (\vec r, t)=
\psi(z,y,t) \xi_0(x,t)$ where $\xi_0$ is the fundamental state of the $x$
direction.  The evolution equation of $\psi$ is then given by
\begin{eqnarray}
 i\hbar\frac{\partial \psi(z,y,t)}{\partial  
t}=-\frac{\hbar^2}{2m}\nabla^2\psi(z,y,t) \nonumber \\
 +\frac{Nu_0}{
L_x}|\psi(z,y,t)|^2\psi(z,y,t),  \label{2d}
\end{eqnarray} 
where $L_x$ denotes the length in the $x$ direction.

As in the
one dimensional case in \cite{villainfpu} we can derive some criteria to
estimate the kind of dynamics the system will have.  We repeat here briefly
these ideas.
We decompose $\psi$ using the basis of the eigenmodes of the two dimensional
box, $\psi(z,y,t) =\sum_{n,m} c_{nm}(t) \phi_n(z) \chi_m(y)$ with 
$\phi_n(z)=
\sqrt{\frac{2}{L_z}} {\rm sin}(\frac{n\pi z}{L_z})$ and 
$\chi_m(y)=
\sqrt{\frac{2}{L_y}} {\rm sin}(\frac{m\pi y}{L_y})$.
Inserting this form into the equation (\ref{2d}) we obtain a set of coupled
differential equations for the coefficients $c_{nm}$.
They can be derived as the Hamiltonian equations of the motion for the complex
degree of freedom $\{c_{nm},i\hbar c^*_{nm} \}$. It is however more
interesting to work in action-angle variables $\{ I_{nm}, \theta_{nm}
\}$ defined  such that $c_{nm}= \sqrt{\frac{I_{nm}}{\hbar}}
e^{i\theta_{nm}}$. The Hamiltonian leading to the differential equations
of motion for the $c_{nm}$ can the be written in action-angle
representation as
\begin{eqnarray}
 H &=&
\sum_{nm}\omega_{nm} I_{nm} \nonumber \\ &+& \frac{\tilde g}{2} \sum_{\vec n
\vec m} V_{\vec n \vec m} (I_{n1m1}I_{n2m2}I_{n3m3}I_{n4m4})^{1/2}
\nonumber \\
 &e&^{-i(\theta_{n1m1}+\theta_{n2m2}-
\theta_{n3m3}-\theta_{n4m4})}, 
\end{eqnarray}  
where $\tilde g= \frac{Nu_0}{L_x \hbar^2}$ and  where \\ 
$$V_{\vec n \vec m}=
\int_0 ^{L_z} dz\ \phi_{n1}\phi_{n2} \phi_{n3}\phi_{n4} \ \int_0^{L_y} dy \
\chi_{m_1} \chi_{m2} \chi_{m3}\chi_{m4}.$$
Here $\vec n$ and $\vec m$ correspond
to  $(n_1, n_2,n_3,n_4)$ and $(m_1,m_2,m_3,m_4)$. We have also
introduced the frequencies $\omega_{nm}$ corresponding to the  mode
labelled $(n,m)$. We have of course $\hbar\omega_{nm}=\frac{n^2 \pi^2
\hbar^2}{2mL_z^2} + \frac{m^2 \pi^2 \hbar^2}{2mL_y^2}= (n^2+16 m^2) \hbar
\omega_{1z}$ where we introduced the fundamental frequency $\omega_{1z}$
in the $z$ direction. 

Compared to the equations
derived in \cite{villainfpu} we have just a doubling of the indices due to the
inclusion of a second dimension. The derivation of the criteria which indicate
the borderline between two different dynamical behaviors is then
straightforward.  The first criterium
 indicates for what conditions one may expect
that the dynamics of the systems will remain close to the linear one. In
order to derive this criterium  we have to compare the frequencies $\omega_{nm}$ of the
linear problem with their first order correction given by the nonlinearities
\cite{villainfpu}. In this case we have 
\begin{equation} 
I_{nm} << \frac{8}{9}
\frac{\omega_{1x}}{\Omega_{int}} \hbar (n^2+ 16m^2) \label{lincon2d}
\end{equation} 
where the natural frequency of the nonlinearities is denoted by
$\Omega_{int}=\frac{N u_0}{\hbar}$.
As the $I_{nm}$'s are normalized to $\hbar$ the high modes of the box will
satisfy this inequality more easily than the low ones for a given value of
the nonlinear parameter $\frac{\Omega_{int}}{\omega_{1z}}$. 

The Chirikov criterium \cite{villainfpu,chirikovreport} on the other hand,
 gives
us an indication about the conditions for which the  dynamics may stay 
regular. This condition follows from the resonance overlap criterion which 
expresses the condition that separatrices associated with two consecutive 
resonances are touching each other. We compare the difference between two 
consecutive bare frequencies with the frequency associated with the first 
order correction. In our case it leads to
\begin{eqnarray}
I_{nm} \leq \frac{16}{9} \frac{\omega_{1z}}{\Omega_{int}}\hbar
(n+16m+\frac{17}{2}).\label{chir2d}
\end{eqnarray}
As the previous inequality, this one is also more easily 
fulfilled  for the highly excited  modes, which
consequently may preserve a regular motion easier than the low ones for an
increasing value of the nonlinearities.
 Clearly for a given nonlinear parameter which is at best of the order of the
typical ratio   of the two energy scale (say less than 30), only the
modes with very low $m$ may not satisfy these inequalities.
As the initial state we choose  is a centered wave packet of widths $L_z/10$ in
the $z$ direction and $L_y/10$ in the $y$ direction without any mean velocity,
we initially populate the modes $(n,m)$ with $n,m=$1 or 3 essentially. This
choice is coherent with what has been done in the one dimensional problem and
is  one of the most favorable in order to see the appearance of 
characteristic nonlinear dynamics.

We consider three cases for which the ratio 
$\frac{\Omega_{int}}{\omega_{1z}}$ takes the values: (a) 4, (b)
10, (c) 101. Experimentally this would correspond to 
condensates whose number of atoms are: (a) few hundreds of atoms, (b) from
few hundreds to 2000, (c) from few thousands to 15000.
As for the one dimensional case, we want to compute the variance and the
statistical entropy associated with a given direction. In the following
paragraph we give the definition of $\sigma_z$ and $S_z$. Exchanging
$z\leftrightarrow y$ in the following expression gives the expressions of
$\sigma_y$ and $S_y$.
In the case of an initially centered wave packet, the preservation of the
symmetry of the wave function simplifies a little bit the  expressions.
We obtain
\begin{eqnarray} 
\sigma^2_z(t)= \int_0^{L_z} dz \ z^2  \Pi(z,t) 
\end{eqnarray}
where $\Pi(z,t)$ is the density of probability of presence at $z$ at the time
$t$,  $\Pi(z,t)= \int_0^{L_y} dy \ |\psi(z,y,t)|^2$.
For the entropy we use  the probability distribution $P(z_n,t) =
\Pi (z_n,t) dz$, giving $S_z(t) = -\sum_k P(z_k,t){\rm
log}P(z_k,t)$, where the $dz$ is just the numerical step used to define
the grid. As before it is more convenient to work with a normalized entropy
$\eta_z$ whose definition is the same as (\ref{entropy}) and with a variance
compared to its value in the uniform case.

\subsection{Numerical results}

We have numerically solved the two dimensional time dependent Gross-Pitaevskii
equation (\ref{2d}) using an explicit method \cite{stenholm}. 
We checked the
conservation of the norm of the wave function (better than $10^{-4}$) and of
the energy (better than 1$\%$). 
In the Figs. 6 and 7 we present the time variations of the variance and the
normalized entropy associated with the different directions for the case (a).
The time is given in unit of $T_1= 2\pi /\omega_{1z}$. This is the reason
why the dynamics in the $y$ direction is presented on a time scale ten
times shorter than for the $z$ direction. As in the one dimensional
case, it is however still long from an experimental point of view. 

\begin{figure}[ht] 
\begin{center}  
\includegraphics[width=6cm]{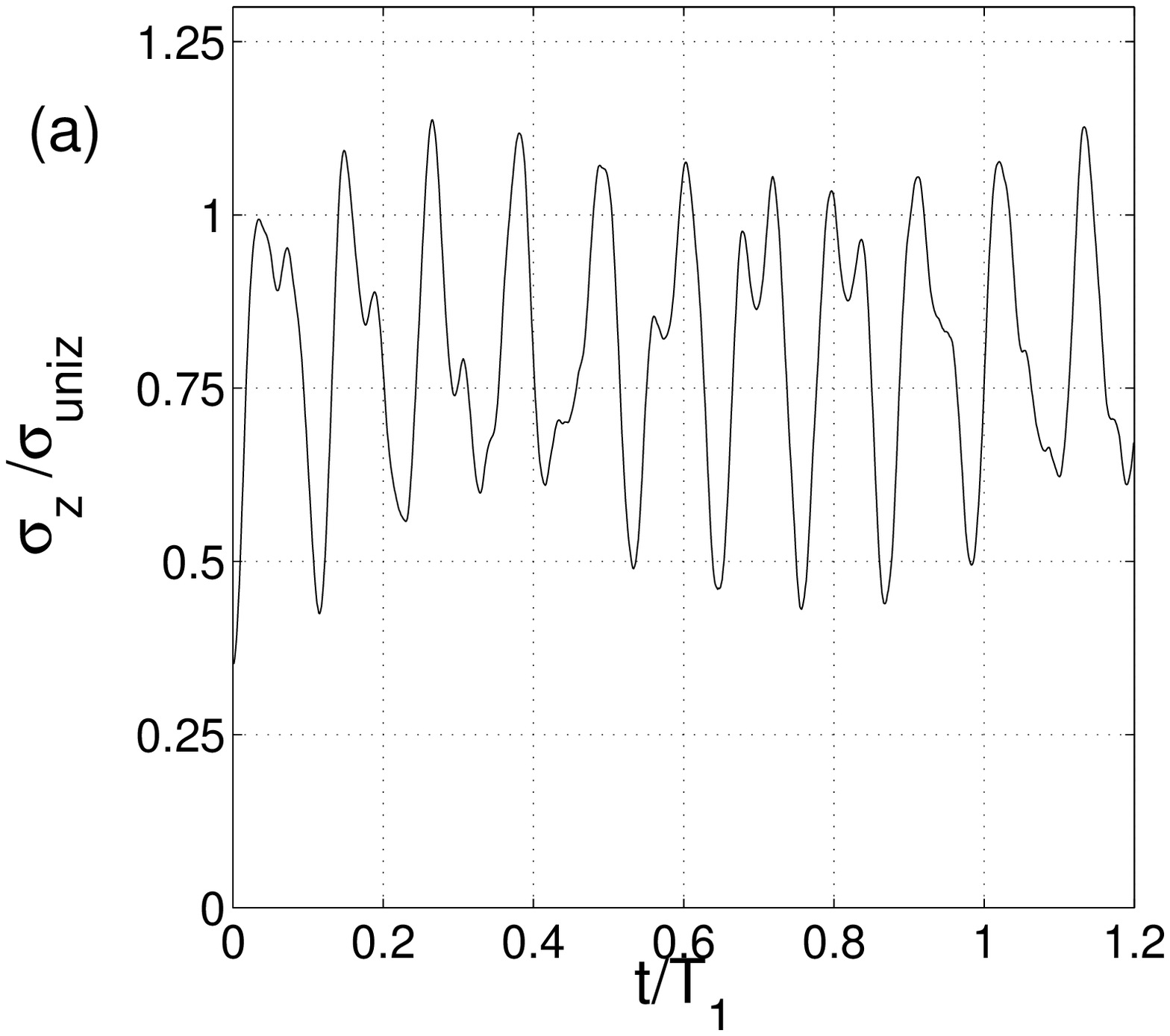} \\
\includegraphics[width=6cm]{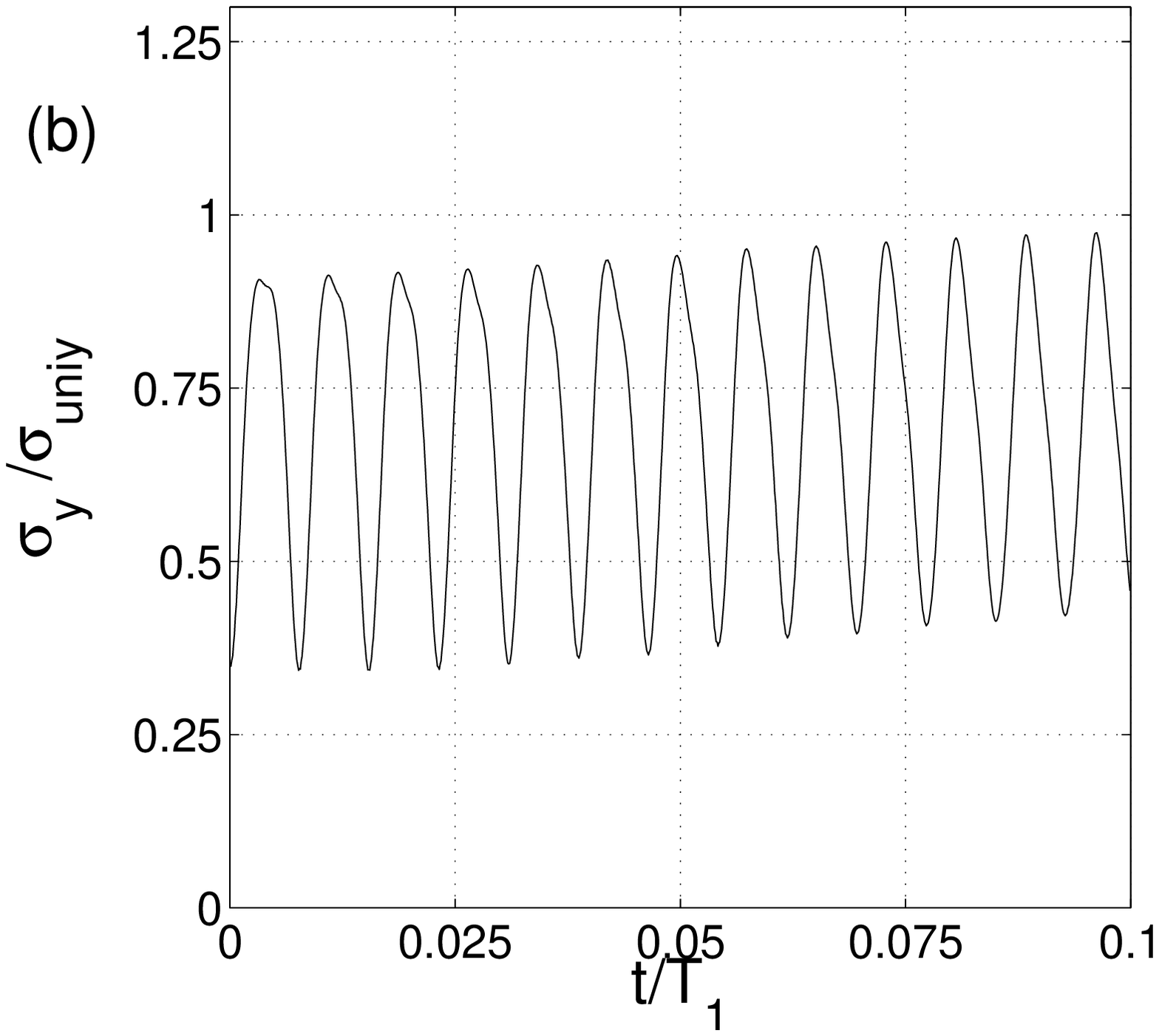}
\label{Fig. 5}
\caption{(a) Temporal variations of $\sigma_z/\sigma_{uniz}$; (b) Temporal
variations of $\sigma_y/\sigma_{uniy}$ for an initial centered state without
mean velocity for  $\frac{\Omega_{int}}{\omega_{1z}}=4$} 
\end{center}   
\end{figure}
\begin{figure}
\begin{center} 
\includegraphics[width=6cm]{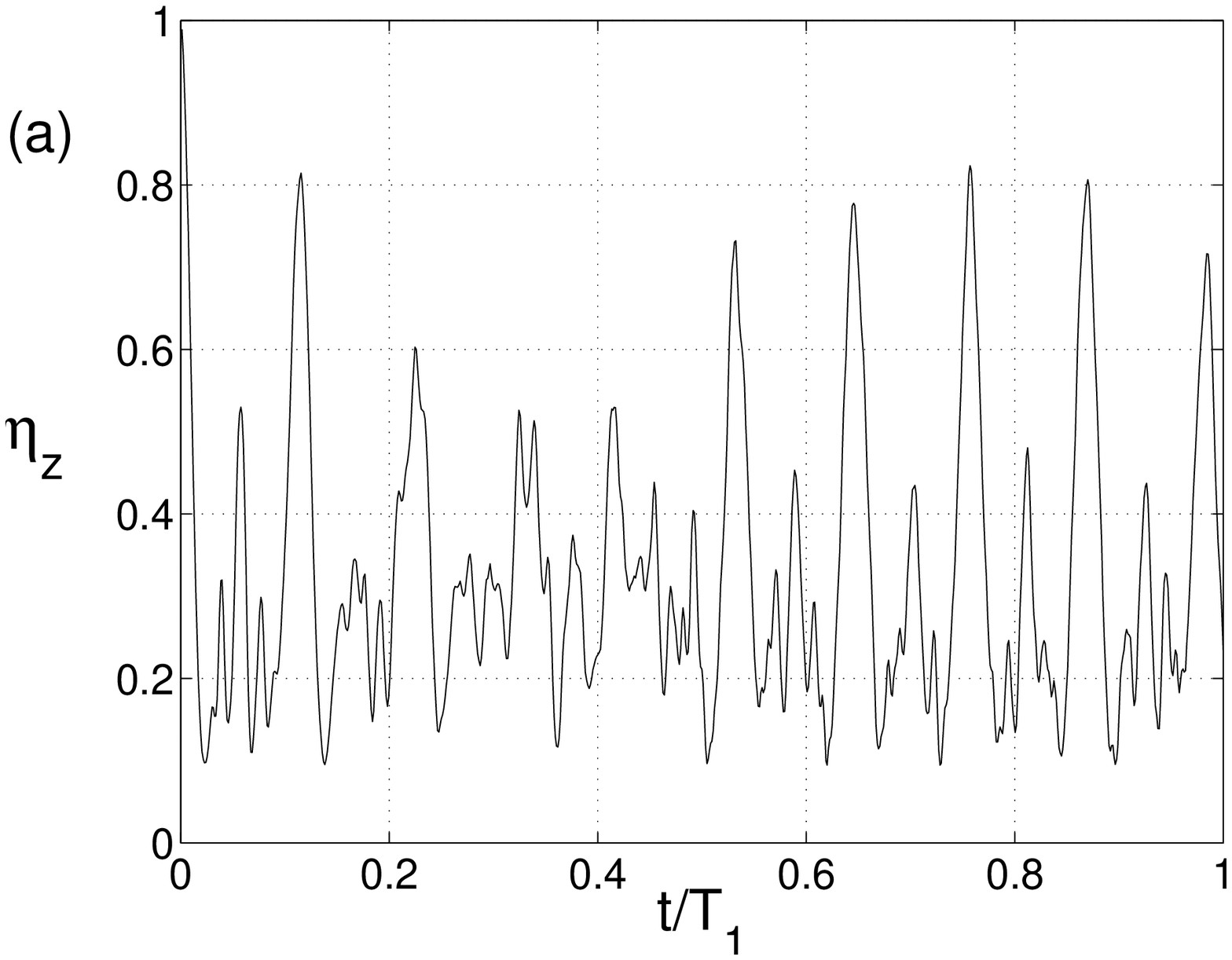}
\includegraphics[width=6cm]{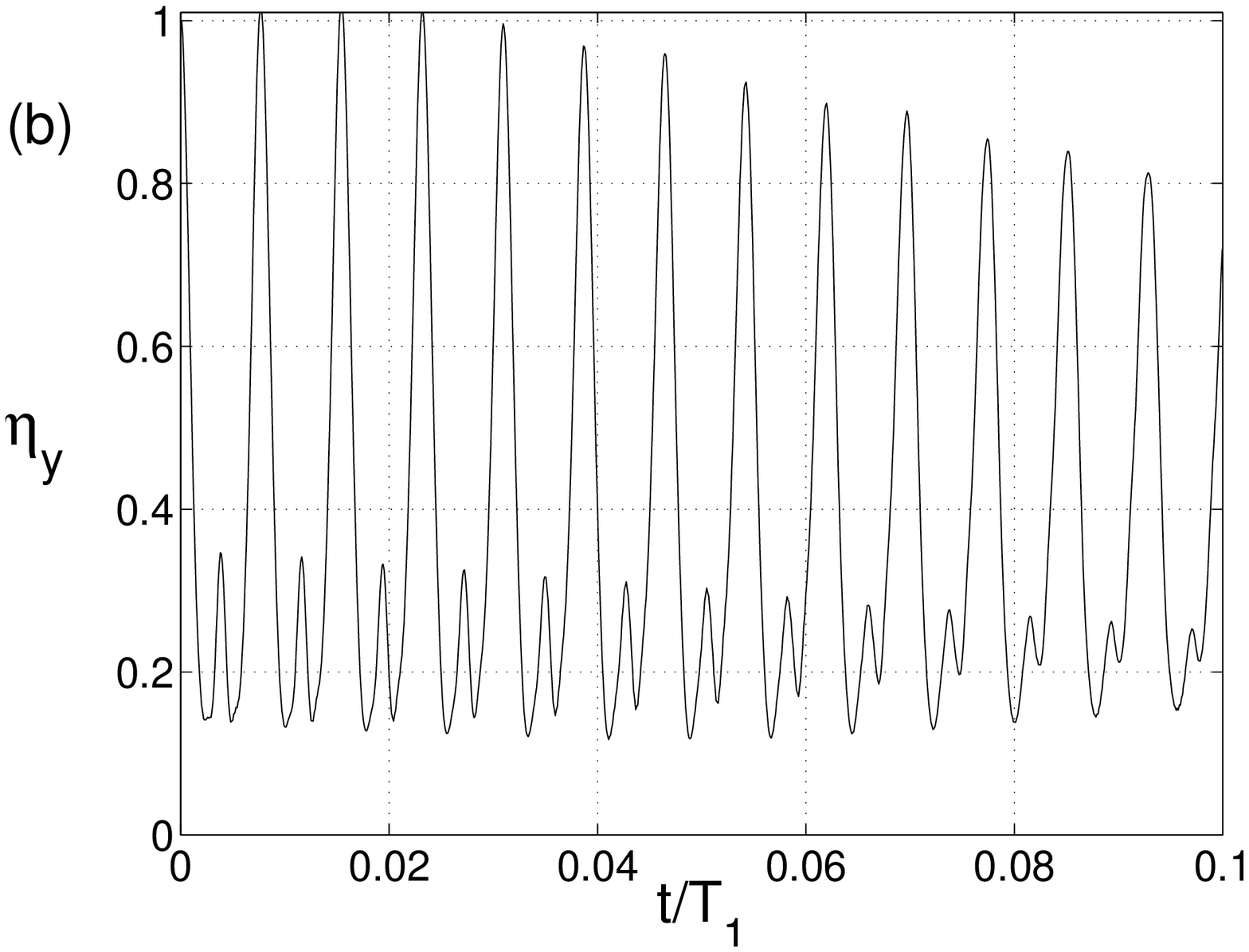}
\label{Fig. 6}
\caption{ time variations  of the normalized entropy associated with the
position for an initially centered wave packet. (a) $\eta_z$, (b) $\eta_y$ for
$\frac{\Omega_{int}}{\omega_{1z}}=4$}  
\end{center}
\end{figure} 

The dynamics is regular as we may expect it for this small value of
nonlinearities. For times shorter than $0.05T_1$, the behavior associated with
the $y$ direction is however closer to a linear one than in the $z$
direction. This appears through the returns of the normalized entropy to the
value 1 as it will be the case for a linear behavior. This means that the $m$
mode distribution did not change during the evolution on the time scale
presented here. For longer times a decrease in the precision of the revivals
appears, but it remains rather small. Indeed, it is only 
visible on the scale of the
normalized entropy although the variance did not exhibit a noticeable
change.  In the $z$ direction
quasi-periodic revivals of the wave packet in this
direction are shown. They correspond  exactly 
to the ones seen in Fig. 2. We can then
conclude that for the smallest value of the nonlinearities considered in the 
one dimensional case, the effects associated with the radial degree of
freedom will not prevent the appearance of nonlinear revivals of the wave
function, and in fact do not play a relevant role.

We turn now to the case of intermediate nonlinearities corresponding to the
case (b).
In Figs. 8 and 9 we present the variations of the variances and the normalized
entropies.
\begin{figure}[ht] 
\begin{center}  
\includegraphics[width=6cm]{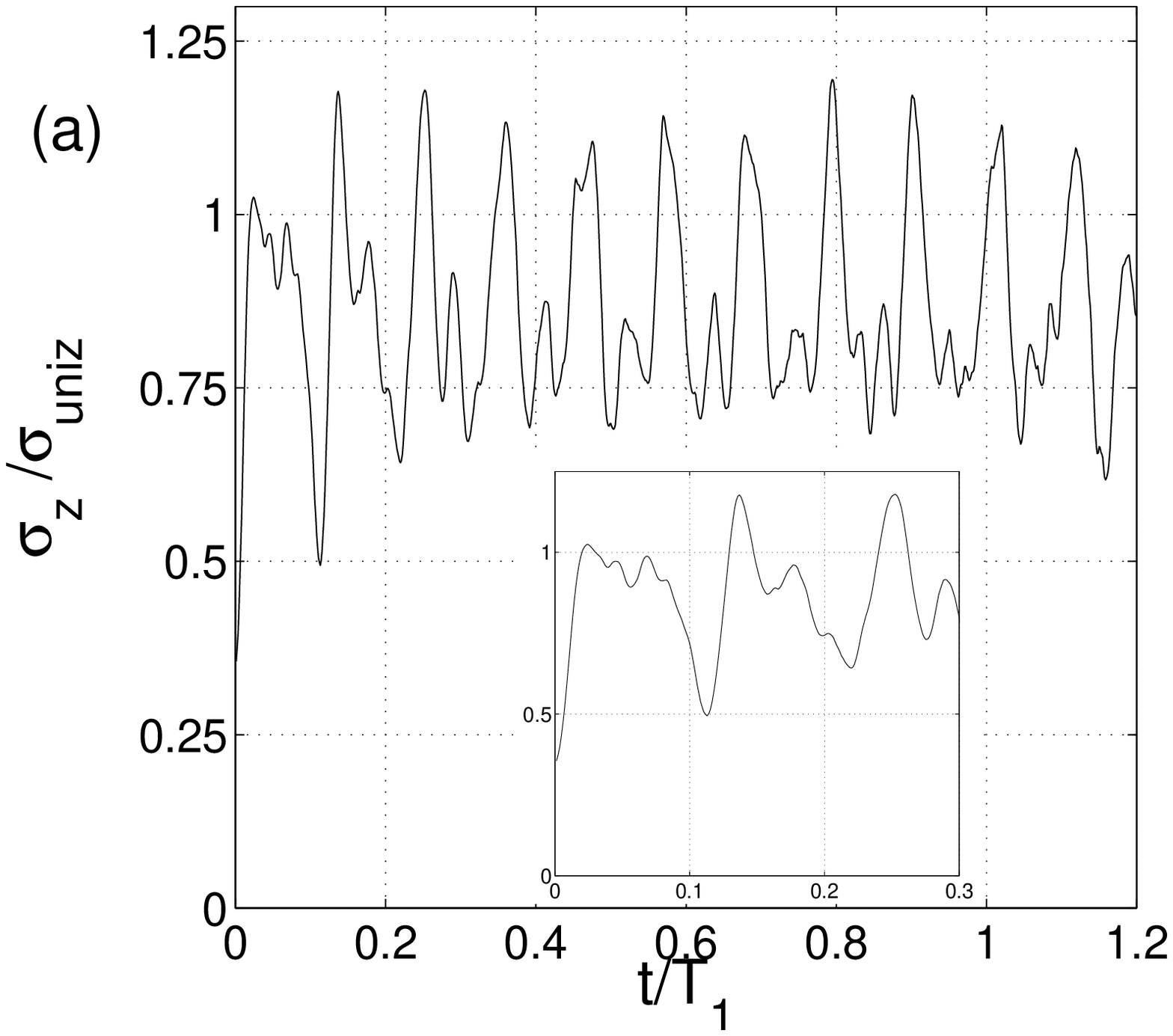} \\
\includegraphics[width=6cm]{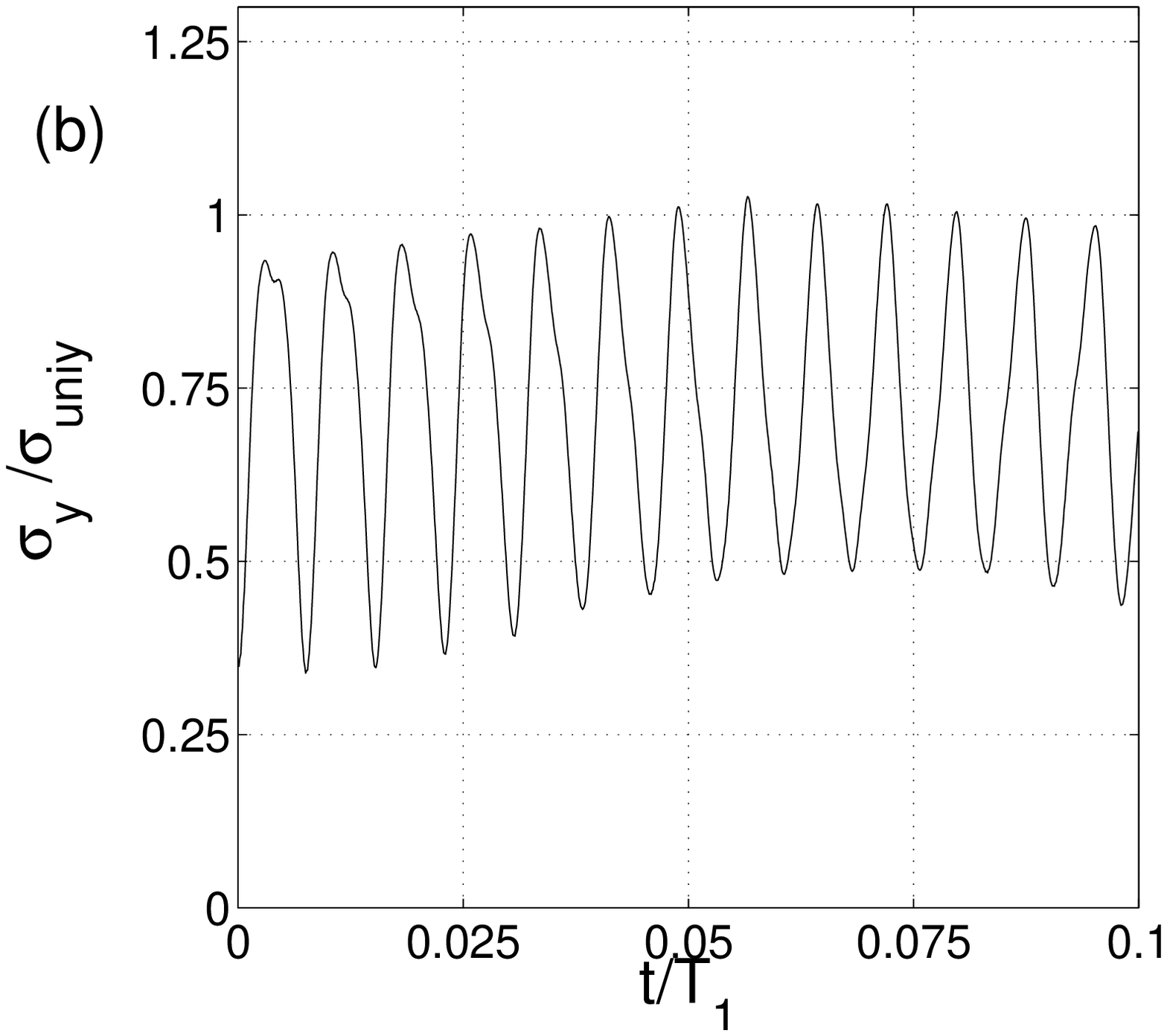}
\label{Fig. 7}
\caption{(a) Temporal variations of $\sigma_z/\sigma_{uniz}$; (b) Temporal
variations of $\sigma_y/\sigma_{uniy}$ for an initial centered state without
mean velocity for  $\frac{\Omega_{int}}{\omega_{1z}}=10$} 
\end{center}   
\end{figure}
\begin{figure}
\begin{center} 
\includegraphics[width=6cm]{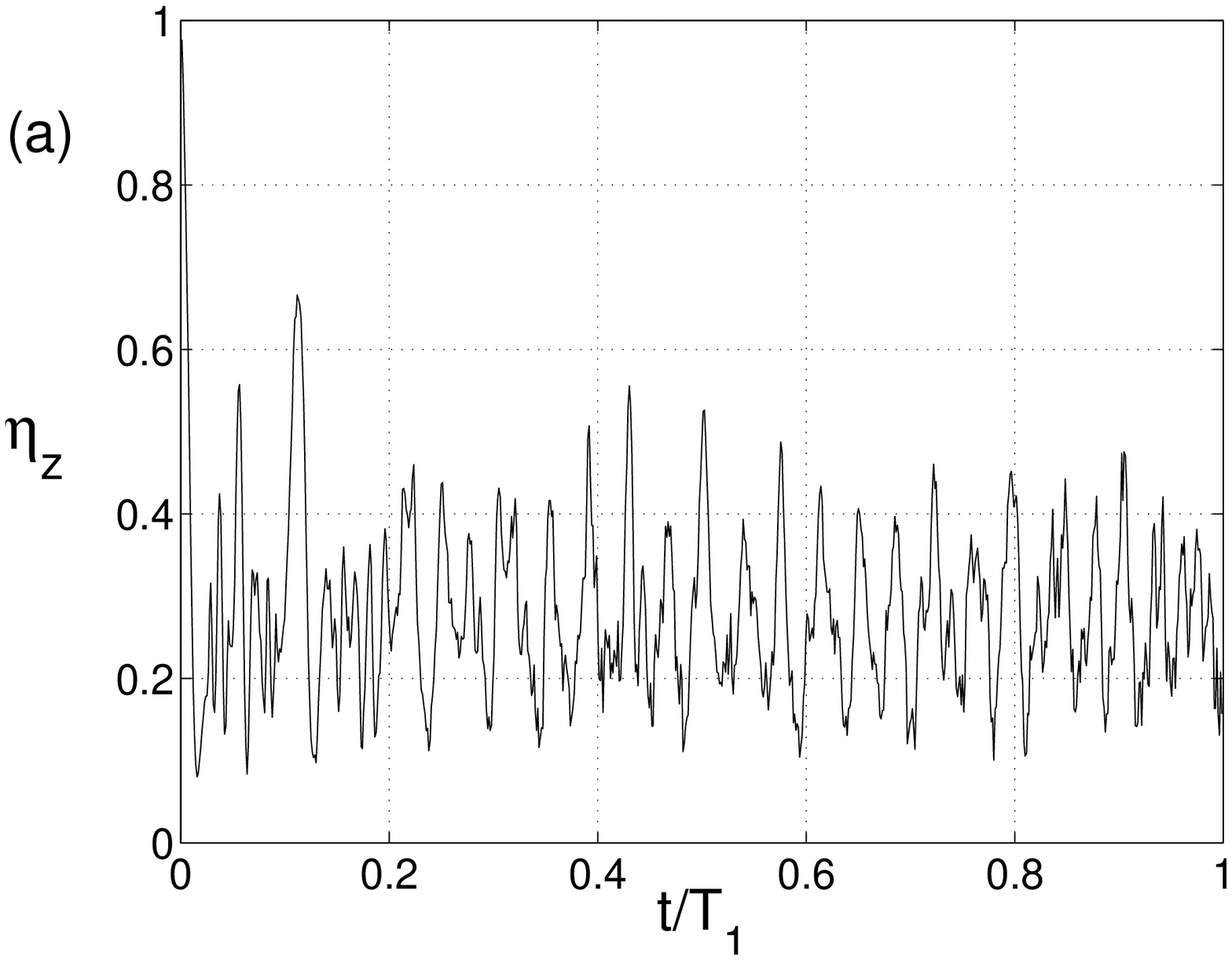}
\includegraphics[width=6cm]{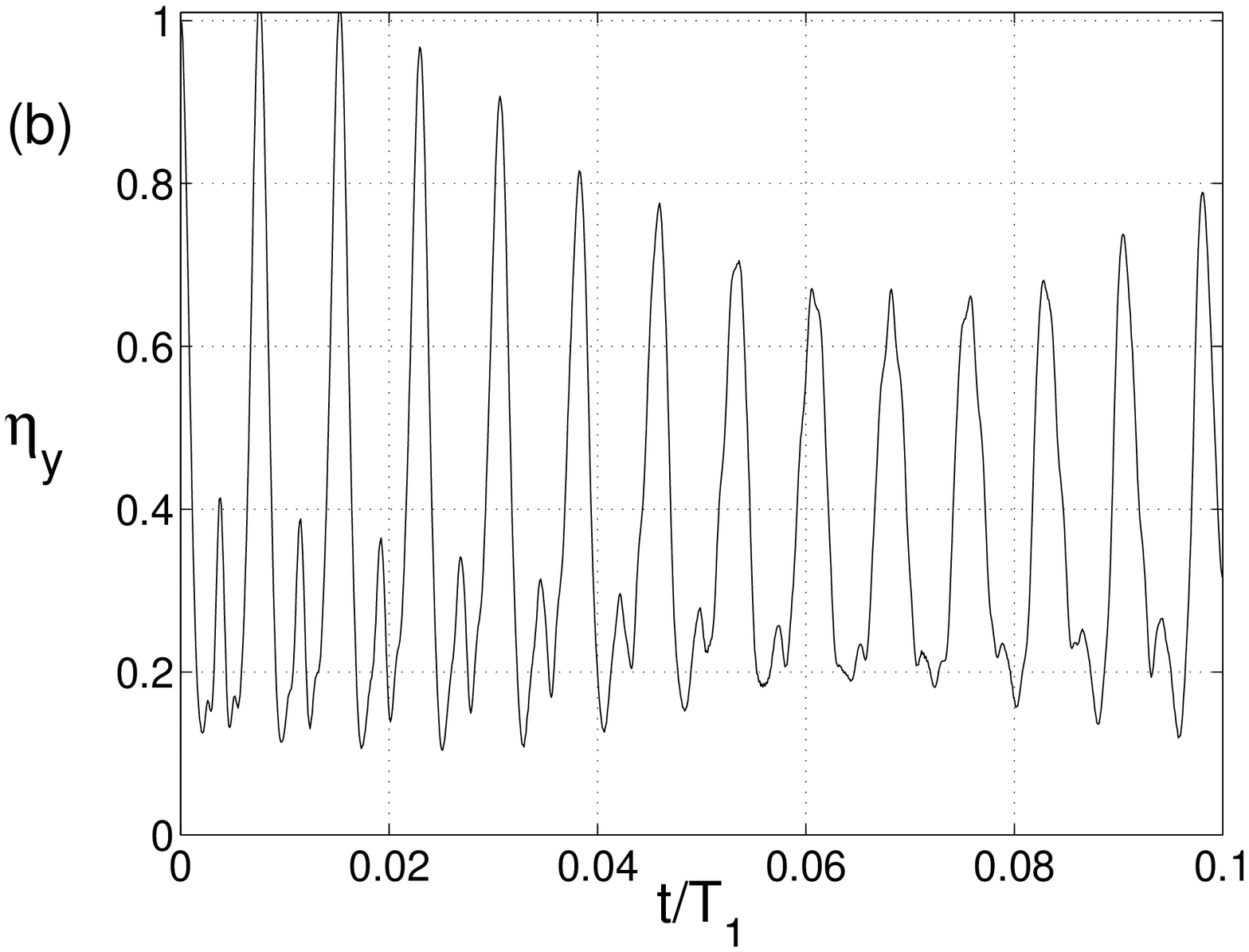}
\label{Fig. 8}
\caption{ time variations  of the normalized entropy associated with the
position for an initially centered wave packet. (a) $\eta_z$, (b) $\eta_y$ for
$\frac{\Omega_{int}}{\omega_{1z}}=10$}  
\end{center}
\end{figure}
 
The dynamics associated with the two directions are completely different.
In the $z$ direction the quasi-periodic spatial relocalizations of the wave
function have vanished. As in the one dimensional case, they have been
replaced by revivals of regular dynamics as can be seen in the inset for the
variance $\sigma_z$. A stochastic dynamics has taken place for this direction.
On the contrary the dynamics in the $y$ direction has remained regular and
quasi-periodic, although compared to the previous case the nonlinearities
begin to influence it stronger. Consequently when the nonlinear
parameter $\frac{\Omega_{int}}{\omega_{1z}}$ 
is comparable to the ratio between 
the energies in {\it z} and {\it y} direction,
revivals of regular dynamics 
in the direction of the small energy scale are expected. The predictions done
in the approximated one dimensional case are then valid for the typical
values of the nonlinearities we have considered (we point out that in this case
we have investigated the behavior for $\frac{\Omega_{int}}{\omega_1}=61$ 
where the
ratio of the scale of energies is of the order of 100).

If we increase more strongly the nonlinearities the coupling between the modes
can no more be neglected for the $y$ direction. This is shown in the Figs.
10 and 11, where we present the variance and the normalized
entropy for the case (c).
\begin{figure}[ht] 
\begin{center}  
\includegraphics[width=6cm]{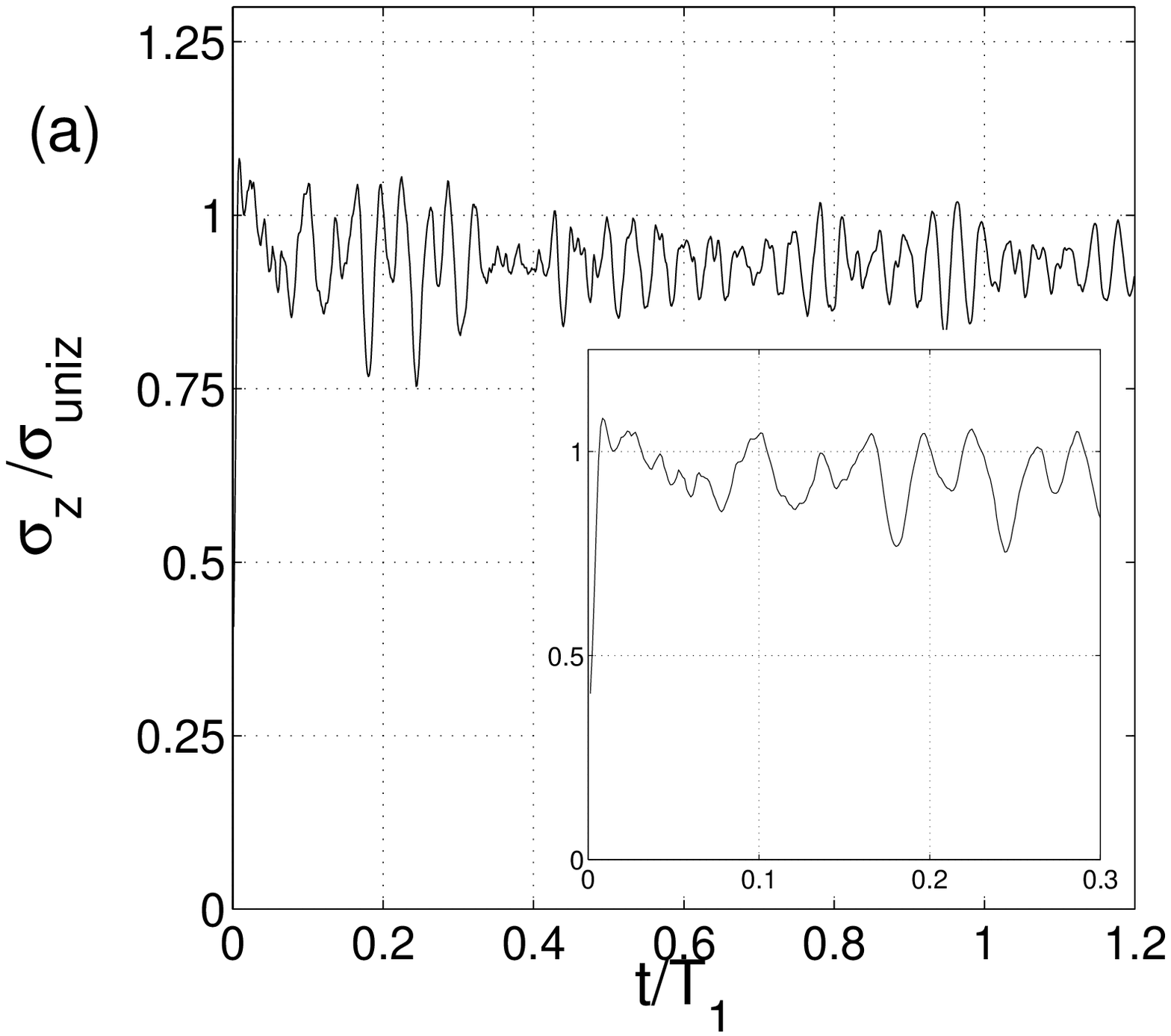} \\
\includegraphics[width=6cm]{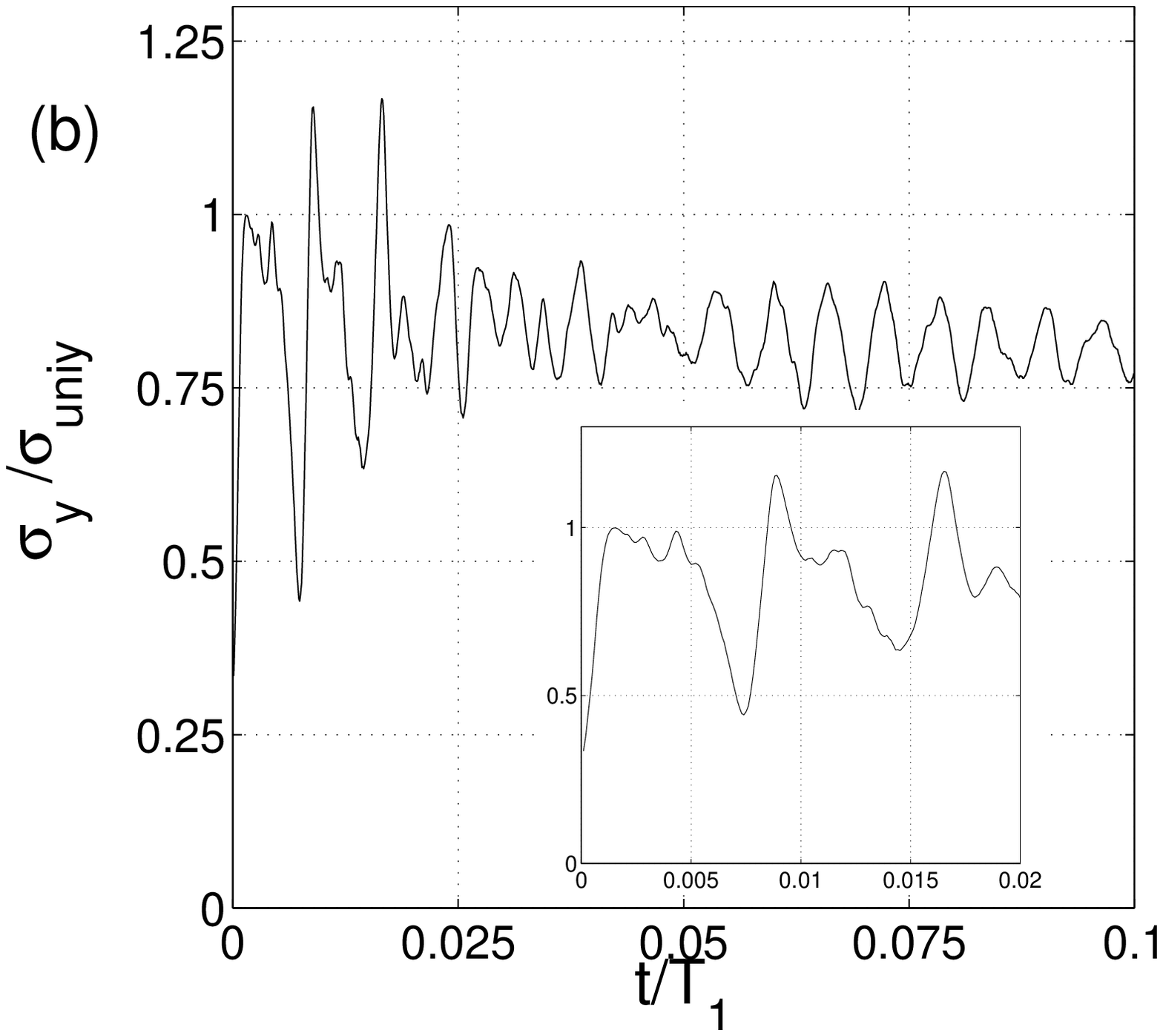}
\label{Fig. 9}
\caption{(a) Temporal variations of $\sigma_z/\sigma_{uniz}$; (b) Temporal
variations of $\sigma_y/\sigma_{uniy}$ for an initial centered state without
mean velocity for  $\frac{\Omega_{int}}{\omega_{1z}}=101$} 
\end{center}   
\end{figure}
\begin{figure}
\begin{center} 
\includegraphics[width=6cm]{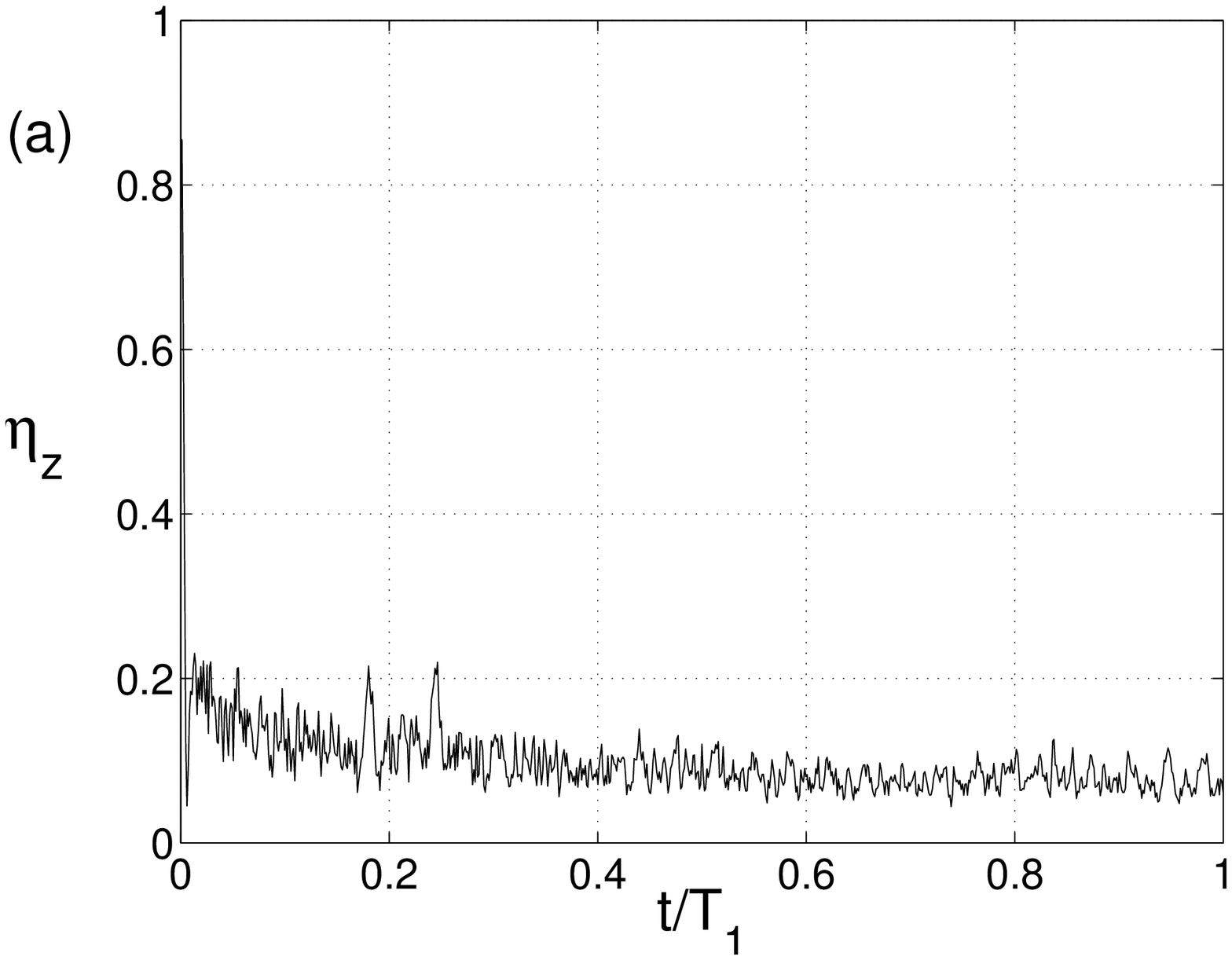}
\includegraphics[width=6cm]{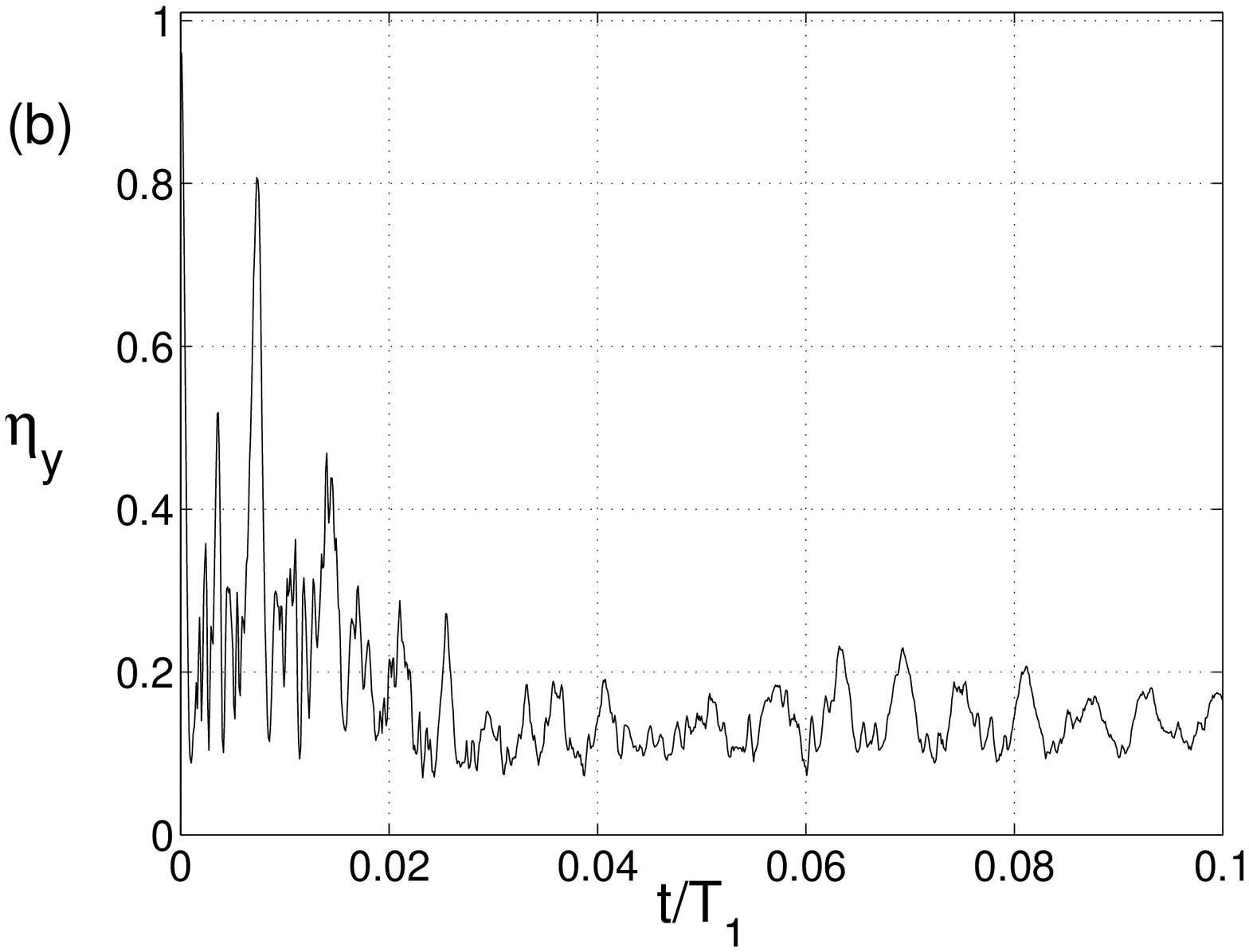}
\caption{Time variations  of the normalized entropy associated with the
position for an initially centered wave packet. (a) $\eta_z$, (b) $\eta_y$ for
$\frac{\Omega_{int}}{\omega_1z}=101$}  
\end{center}
\end{figure} 

The very fast decrease of the two normalized entropies indicates a strong
relaxation of the system towards a nearly uniform state.
The variance in the $z$ direction reaches almost a steady state. It oscillates
around its value corresponding to the uniform case and the revivals of regular
dynamics of the previous case are still visible, but vanish very
quickly. In the $y$
direction the situation is similar. The previous revivals of the wave packet
in this direction have been replaced by revivals of regular dynamics, whose
amplitudes decrease quickly with time. The dynamics have developed
stochasticity as may be expected from the application of the Chirikov
criterium (\ref{chir2d}) for this value of the nonlinear parameter.

\section{Conclusions}

In this paper we have studied the nonlinear dynamics of a 
condensate trapped in an optical box like potential. The scheme 
we have in mind is a condensate located in 
the center of a box-like potential where we monitor the time evolution of 
the wave packet by looking at the associated entropy and variance in order to
detect the main  properties of the dynamics. The behavior of these functions
allows to confirm the predictions done previously in \cite{villainfpu}, namely
that for small nonlinearities (corresponding to situations with few
thousands of atoms at most) the system will exhibit nonlinear revivals 
through quasi-periodic spatial  relocalizations. On the other hand, 
for stronger
nonlinearities (corresponding typically to a number of atoms of few  tens of
thousands ) the system will follow an effective relaxation dynamics, for which
the wave function revivals are replaced by revivals of regular dynamics. The
observed revival times in a perfect box potential are shown to be
rather too long  to be experimentally measurable.
Fortunately using a more realistic trapping potential with a pair of blue
detuned lasers as ``end caps'' for the doughnut mode  laser configuration,
there is a substantial reduction of the revival times,  which can then 
be shorter
than the typical condensate life time. It is important to note that the 
present study demonstrates  the possibility of observation 
of the revivals of a mesoscopic object, which is interesting from
an experimental point of view. It shows, however, 
 also irreversibility in a closed system which
is interesting from a more fundamental point of view.

In the numerical treatment we have first assumed all 
the dynamics was taking place
in one dimension. In other words the transversal dynamics of the condensate 
was 
frozen. This assumption, based on an effective strong radial confinement, is
valid as long as the (nonlinear) coupling is not too strong. In order to 
justify
the one dimensional treatment and to study the  break-down of it, we have
performed a two dimensional simulation using a  perfect box potential. By
choosing a trap aspect ratio significantly lesser  than one, we have 
demonstrated  that
the coupling between the two dimensions is present only at large values of
the parameter $\Omega_{int}/\omega_{1z}$ describing  the nonlinear interaction.
Hence,  the one dimensional study provides 
 indeed  a  good approach for the physical
parameters used here. The coupling between  the two dimensions is mainly
governed by the different energy scales of the  eigenmodes in {\it z} and {\it
y} direction due to the aspect ratio.  

Finally, we would like to point out that 
very recent experimental progress of the Hannover group allows to
expect that the observation of the condensate wave packet and transition
to irregularity will be feasible within the times of the order of 1s 
\cite{hannoverlaser}.

\acknowledgements

We are grateful to A. Sanpera, K. Sengstock, G. Birkl and W. Ertmer for 
helpful discussions. This work has been supported by Deutsche
Forschungsgesellschaft (SFB407), the TMR network ERBXTCT96-0002 and the ESF
PESC Program BEC2000.

\end{document}